\documentclass[useAMS,usenatbib]{mn2e}

\usepackage{graphicx}
\usepackage{rotating}
\usepackage{amsmath}
\usepackage{amssymb}
\usepackage{lscape}
\usepackage{natbib}
\usepackage{keyval}
\usepackage{lastpage}
\usepackage{array}
\usepackage{dcolumn}
\usepackage{comment}
\usepackage{pifont}
\usepackage{natbib,times}
\usepackage[T1]{fontenc}
\usepackage{aecompl} 

\newcommand{\tbgscold}{\hbox{$T_\mathrm{C}^{\,\mathrm{ISM}}$}}

\newcommand{\ldust}{\hbox{$L_{\mathrm{d}}$}}

\newcommand{\mdms}{\hbox{$M_{\mathrm d}/M_\ast$}}

\newcommand{\md}{\hbox{$M_{\mathrm d}$}}
\newcommand{\mstar}{\hbox{${\mathrm M}_\ast$}}
\newcommand{\lco}{\hbox{$L^{\prime}_{\rm{CO(1-0)}}$}}
\newcommand{\msun}{\hbox{$\rm{M_{\odot}}$}}

\newcommand{\lfir}{\hbox{$L_{\rm{FIR}}$}}
\newcommand{\siggas}{\hbox{$\Sigma_{\mathrm {H}_2}$}}
\newcommand{\sigsfr}{\hbox{$\Sigma_{\mathrm {SFR}}$}}
\newcommand{\alphaco}{\hbox{$\alpha_{\rm{CO}}$}}

\def\arcsec{\hbox{$^{\prime\prime}$}}

\def\deg{\hbox{$^\circ$}}

\def \be {\begin{equation}}
\def \ee {\end{equation}}

\def\gsim{\mathrel{\lower0.6ex\hbox{$\buildrel {\textstyle >}
 \over {\scriptstyle \sim}$}}}
\def\lsim{\mathrel{\lower0.6ex\hbox{$\buildrel {\textstyle <}
 \over {\scriptstyle \sim}$}}}

\def \halpha {H$\alpha$}
\def \nii {[N~{\sc ii}]}
\def \oiii {[O~{\sc iii}]}

\def \hb {H$\beta$}

\def \mustar {$\mu^*$}

\title[Evolution of the cold ISM  after a starburst]{The evolution of the cold interstellar medium in galaxies following a starburst\thanks{{\it Herschel} is an ESA space observatory with science instruments provided by European-led Principal Investigator consortia and with important participation from NASA.}}

\author[K. Rowlands et al.]
{K. Rowlands$^1$\thanks{E-mail:ker7@st-andrews.ac.uk}, V. Wild$^1$, N. Nesvadba$^2$, B. Sibthorpe$^3$, A. Mortier$^4$, M. Lehnert$^5$, 
\newauthor E. da Cunha$^6$
\\ 
$^1$(SUPA) School of Physics \& Astronomy, University of St Andrews, North Haugh, St Andrews, Fife, KY16 9SS, UK \\ 
$^2$ Institut d'Astrophysique Spatiale, CNRS, Universit\'{e} Paris-Sud, Bat. 120-121, F-91405 Orsay, France \\
$^3$ SRON Netherlands Institute for Space Research, Zernike Building, P.O. Box 800, 9700 AV Groningen, The Netherlands \\
$^4$ School of Physics \&\ Astronomy, The University of Nottingham, University Park Campus, Nottingham, NG7 2RD, UK \\
$^5$ Institut d'Astrophysique de Paris, UMR 7095, CNRS, Universit\'{e} Pierre et Marie Curie, 98 bis boulevard Arago, 75014 Paris, France \\
$^6$ Max Planck Institute for Astronomy, Konigstuhl 17, 69117, Heidelberg, Germany \\
}

\begin{document}
\maketitle
\begin{abstract}
We present the evolution of dust and molecular gas properties in a sample of 11 $z\sim0.03$ starburst to post-starburst (PSB) galaxies selected to span an age sequence from ongoing starburst to 1\,Gyr after the starburst ended. All PSBs harbour significant molecular gas and dust reservoirs and residual star formation, indicating that complete quenching of the starburst due to exhaustion or expulsion of gas has not occurred during this timespan. As the starburst ages, we observe a clear decrease in the star-formation efficiency, molecular gas and SFR surface density, and effective dust temperature, from levels coincident with starburst galaxies to those of normal star-forming galaxies. These trends are consistent with a natural decrease in the SFR following consumption of molecular gas by the starburst, and corresponding decrease in the interstellar radiation field strength as the starburst ages. The gas and dust contents of the PSBs are coincident with those of star-forming galaxies and molecular gas-rich early-type galaxies, and are not consistent with galaxies on the red-sequence. We find no evidence that the global gas reservoir is expelled by stellar winds or AGN feedback. Our results show that although a strong starburst in a low-redshift galaxy may cause the galaxy to ultimately have a lower specific SFR and be of an earlier morphological type, the galaxy will remain in the ``green valley'' for an extended time. Multiple such episodes may be needed to complete migration of the galaxy from the blue- to red-sequence. 
\end{abstract}


\section{Introduction}

It has long been known that the galaxy population displays a colour bimodality
\citep{Strateva01, Blanton03, Baldry04, Bell04b}. The dearth of green valley galaxies relative to the optically blue and red
populations, together with the gradual build up of mass on the red sequence over cosmic time, has led to the idea that many galaxies must undergo a rapid change in colour due to quenching of star-formation on a $\lesssim1$\,Gyr timescale \citep{Martin07, Kaviraj07, Schawinski07, Wild09, Yesuf14}.

Post-starburst (PSB) galaxies are candidates for such a transition population. These galaxies have undergone a starburst in the recent ($\lesssim 1$\,Gyr) past which has since declined rapidly in strength, leaving a dominant A/F-star population. Local PSBs which have completely quenched their star formation are found to have predominantly early-type morphologies \citep{Wong12, Mendel13}, 
and a large fraction exhibit morphological disturbances (\citealp{Zabludoff96, Blake04, Goto05, Yang08, Pracy09}, Pawlik et~al. in prep). This has led to the hypothesis that the starburst was triggered by a merger or interaction, at least in the local Universe where gas-to-stellar mass ratios of massive galaxies are relatively low \citep[][but see \citet{Dressler13} for an alternative view]{Bekki01, Bekki05, Snyder11}. Simulations show how an interaction may funnel gas towards the centre of the galaxy \citep[e.g.][]{Mihos_Hernquist94, Mihos_Hernquist96, Barnes_Hernquist96}, leading to an increase in gas density and subsequent strong star formation, ultimately consuming the entire gas reservoir. 

While the number density of PSB galaxies at $z\gtrsim1$ indicates that they could represent a significant channel for red sequence growth \citep{Wild09, Whitaker12a, Yesuf14}, the true global importance of PSB galaxies for the wider picture of galaxy evolution depends on whether the quenching of star formation is permanent or temporary.

To reach the red sequence, the gas reservoir in galaxies must be exhausted, and/or prevented from forming stars via feedback mechanisms \citep[e.g.][]{Benson03, DiMatteo05}. Using simple energetics arguments, \citet{Kaviraj07} suggested  that supernova feedback is consistent with being the dominant quenching mechanism of star formation in nearby low stellar mass ($\mstar<10^{10}$M$_{\odot}$) PSBs, and feedback from active galactic nuclei (AGN) dominates in high mass ($\mstar>10^{10}$M$_{\odot}$) PSBs (see also \citealt{Wong12}). Smoothed-particle hydrodynamic simulations of merging galaxies routinely show how the starburst initially declines through the depletion of gas supplies, but additional energy must be injected into the interstellar medium to completely quench star formation throughout the entire galaxy. Energy input from an AGN is usually invoked to complete the transition of galaxies from blue to red sequence, through gas heating or expulsion \citep[e.g.][]{Springel05, Hopkins07, Khalatyan08, Kaviraj11}. However, there remains no direct observational evidence for such a scenario (although see \citealt{Alatalo14}).

Due to the rarity of PSBs in the local Universe, they are not routinely included in large neutral and molecular gas surveys. Without the need for selection with moderate signal-to-noise optical spectroscopy, close pair/pre-merger samples and IR bright starburst/coalescence phases have been better studied with targeted observations \citep[e.g.][]{Sanders91, Braine_Combes93, Solomon97, Casasola04, Boquien11a, Garcia-Burillo12}. Recently, \citet{Zwaan13} detected HI gas in 6/11 galaxies with PSB populations at $z\sim0.03$, finding that they have atomic gas-to-stellar mass ratios between those of early-type galaxies (ETGs) and spirals. Although these results show that local PSBs may not have run out of gas entirely, HI is not directly connected to star formation as it dominates the interstellar medium (ISM) when gas surface densities are $<10$\,M$_\odot$pc$^{-2}$ \citep{Bigiel08}. Knowledge of their molecular gas contents is required to ascertain why galaxies have stopped forming stars. It is also advantageous to study PSBs with a range of ages, and in conjunction with samples of starbursts from which they originated, so we can track how the state of the ISM alters as the impact of the starburst decays. This has not been the case in previous studies, where stringent limits placed on nebula emission line strengths ensures that all star formation has completely ceased and therefore only the oldest PSBs (and those without Type-II AGN) are selected for follow-up. 

In this paper we examine the cold gas and dust properties of a sample of starbursts and PSBs at $z\sim0.03$ with a range of ages from 0 to 1\,Gyr after the starburst, selected from the spectroscopic sample of the Sloan Digital Sky Survey (SDSS) Data Release 7 \citep[DR7, ][]{SDSS_DR7}. 
We focus purely on galaxies with significant bulges (selected with high stellar surface mass density, \mustar) where the star formation processes are more extreme than in disc dominated systems \citep{Kennicutt98}, and the majority of accretion onto black holes occurs \citep{Kauffmann03b}. The sample is selected purely based on the properties of the stellar population of the galaxies, i.e. with no cut on nebular emission line strength. This serves two purposes: (1) it does not bias against objects containing obscured (narrow line) AGN which may be more prevalent in PSBs than other phases of galaxy evolution \citep{Yan06, Georgakakis08, Brown09, Yesuf14}, and (2) allows us to select PSBs with a range of ages, as starbursts are not instantaneous, but rather decay over $\sim100$\,Myr timescales meaning nebular emission lines are expected to be present even during the PSB phase \citep{Yan06, Falkenberg09, Snyder11}. By selecting purely based on known physical properties of the stellar population, (i.e. starburst age and burst strength), we can select a complete sample of objects in a physically meaningful sense.  
In Section \ref{sec:sample} we describe the method used to select a
complete sample of starburst to post-starburst galaxies. In Section~\ref{sec:data} 
we describe the multi-wavelength data obtained for this study and our spectral energy distribution (SED) fitting method. The results are presented in
Section~\ref{sec:results} and our conclusions in Section~\ref{sec:conc}. We adopt a cosmology with $\Omega_m=0.30,\,\Omega_{\Lambda}=0.70$ and $H_o=70\, \rm{km\,s^{-1}\,Mpc^{-1}}$.

\section{Sample Selection}
\label{sec:sample}

\begin{table*}
  \caption{\label{tab:obsns} Object name, position, SDSS DR7 specobjid, optical redshift
    and total on-source integration time for the millimetre observations. }
\vspace{0.1cm}
  \begin{tabular}{lccccc} \hline
    Object ID & RA & Dec & specobjid & z & t$_{\rm int}$ \\ 
              & (J2000 deg) & (J2000 deg) & & & (min) \\ \hline
PSB1  &233.132&57.8829&173331940055711744&0.039 &570 (600)$^a$   \\
PSB2  &228.951&20.0224&607094233467191296&0.036 &240             \\
PSB3  &225.401&16.7297&782453208889950208&0.032 &300             \\
PSB4  &246.455&40.3452&330115272746729472&0.029 &240             \\
PSB5  &244.398&14.0523&711521640860614656&0.034 &240             \\
PSB6  &252.924&41.6684&177834388912865280&0.043 &120             \\
PSB7  &249.495&13.8594&622289631202770944&0.047 &240             \\
PSB8 &232.702&55.3288&173055146299752448&0.046 &420             \\
PSB9 &239.568&52.4893&174175084368363520&0.049 &480             \\
PSB10 &247.179&22.3971&442707057922015232&0.034 &480             \\
PSB11   &237.803&14.6964&709269844574339072&0.048 &360             \\
\hline
  \end{tabular}
  \begin{minipage}{\textwidth}
    $a$: Combined spectra contain 38 scans at CO(1-0) and 40 scans at
    CO(2-1).
  \end{minipage}
\end{table*}

\begin{figure}
\includegraphics[width=0.48\textwidth]{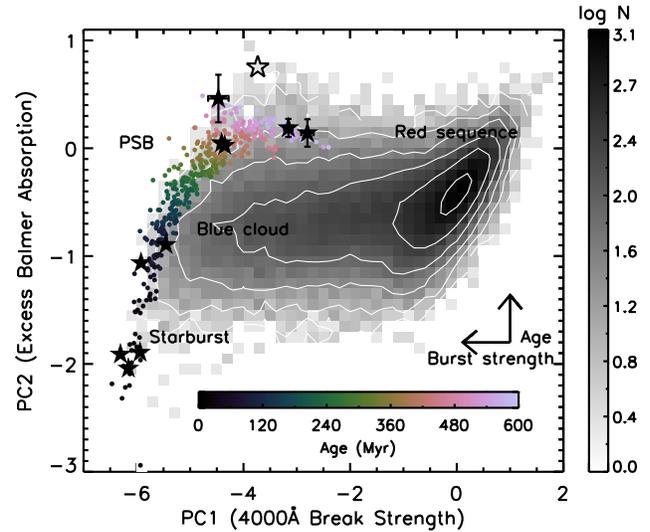}
\caption{The 4000\AA\ break strength and Balmer absorption line
  strength as measured by a principal component analysis of the 4000\AA\  spectral region of the targeted galaxies (stars). The greyscale indicates the distribution of all SDSS DR7 galaxies with spectral
  per-pixel-SNR$>8$ in the $g$-band, redshift $0.01<z<0.07$ and
  stellar surface mass densities $>3\times10^8{\rm
    M_\odot}$\,kpc$^{-2}$. The coloured dots indicate the (post-)starburst sample
  from which the targets were selected. The open star indicates the
  position of the additional target `PSB11', an older PSB
  galaxy than the completeness limit of the original starburst sample. 
  Errors are indicated only where they are significantly larger than the
  symbol size.}\label{fig:pc12}
\end{figure}

In the integrated optical fibre spectrum of a galaxy the different
signatures of stars of different ages can be used to obtain
information about a galaxy's recent star-formation history (SFH). To define
our sample we make use of two particular features of optical spectra:
the 4000\AA\ break strength and Balmer absorption line strength. In
Fig.~\ref{fig:pc12} we show the distribution of two spectral indices
which parametrise these features for 70,000 bulge-dominated galaxies with stellar surface mass density \mustar$= {\mathrm M}_\ast/(2 \pi r_{50,z}^2) > 3\times10^8$\msun\,kpc$^{-2}$ at $0.01<z<0.07$ in the SDSS. These two spectral indices are based on a Principal Component Analysis (PCA) of the 3175--4150\AA\ region of the spectra, and describe the strength of the 4000\AA\ break, and excess
Balmer absorption over that expected for the 4000\AA\ break strength
\citep{Wild07}. As expected for bulge-dominated galaxies, the majority of the galaxies show no evidence of
recent or current star formation, they form the ``red-sequence'' which
lies on the right, with strong 4000\AA\ break strength from the old
stars.  Galaxies that are forming stars have younger mean stellar ages
and therefore weaker 4000\AA\ break strengths and form the
``blue-sequence''. A small number of galaxies are undergoing a
``starburst'' i.e. there has been a sharp increase in the galaxy's
star-formation rate (SFR) over a short timescale ($\sim10^7$ years).  These
galaxies are identified by their unusually weak Balmer absorption
lines, strong UV--blue continua, and weak 4000\AA\ breaks i.e. spectra
dominated by light from O/B stars. These objects lie in the lower left
of Fig. \ref{fig:pc12}. As the starburst ages to a few $10^8$yrs,
the Balmer absorption lines increase in strength as the galaxy passes
into the PSB phase
\citep{DresslerGunn83, CouchSharples87} i.e. A/F star light
dominates the integrated galaxy spectrum. These objects lie in the
upper left of Figure~\ref{fig:pc12}. A comparison of our selection technique to the classical definition of PSB galaxies (e.g. lack of H$\alpha$, [OII] and deep Balmer absorption) is presented in Appendix~\ref{sec:PSB_classical_selection}.

\subsection{Measuring starburst age}
The rapidly changing appearance of the galaxy optical spectrum during and following a starburst, together with the enhanced luminosity of young stars compared to the old (non-starburst) population, allows the robust measurement of starburst age. A full Bayesian spectral synthesis model analysis of the optical spectra, allowing for variations in dust attenuation and previous SFH, yields uncertainties of order 20\,Myr for the youngest objects to 100\,Myr as the starburst ages to $\sim$1\,Gyr \citep{Wild10}. Unfortunately, burst strength at a given starburst age is less well constrained for individual objects, and is more dependent on the underlying SFH of the galaxy. At older ages than $\sim$600\,Myr a degeneracy becomes apparent between burst age and mass, leading \citet{Wild10} to restrict their evolutionary sequence to ages younger than this. 

Through comparison with population synthesis models, using simple toy model star formation histories or more complex histories derived from
simulations, \citet{Wild07, Wild09} showed that the shape of the left
hand side of the distribution in Fig. \ref{fig:pc12} describes the
evolutionary track of a starburst galaxy, with time since the
starburst increasing from bottom to top, and burst strength increasing
from right to left. The galaxies lying at the outermost edge of the distribution have undergone the strongest recent bursts of star formation in the entire sample. At these low-redshifts, these starbursts are not strong; Bayesian fits to spectral synthesis models imply typical burst mass fractions (i.e. fraction of stellar mass formed in the burst) of $\sim$10\%. The models show that if galaxies that had undergone stronger bursts existed, they would lie to the left of the distribution at intermediate starburst ages, where no galaxies are observed. Thus the visualisation of the full distribution of objects, rather than fits to individual objects, allows us to better constrain the parameters of the population as a whole.  

The coloured dots in Fig. \ref{fig:pc12} are selected to form a complete (constant number per unit starburst age) sample of the 400 strongest starbursts in SDSS galaxies with high \mustar. The definition of this sample is discussed in greater detail in \citep[][hereafter WHC10]{Wild10}. Briefly, a Bayesian fit was performed to the PCA indices of the individual galaxies to determine starburst age (${\rm t_{SB}}$) using a set of $10^7$ spectral synthesis models with star formation histories composed of an old bulge population with superposed exponentially decaying starbursts. Burst mass fraction, burst age, decay rate and dust content were free to vary with uniform priors. The resulting errors on the starburst age therefore take into account the potential variations in recent SFH and dust content of the galaxies and account for possible degeneracies between model parameters.

\subsection{Sample for follow-up millimetre and far-infrared observations}
From this base sample of galaxies which have undergone the strongest recent bursts of star formation in the local Universe, we selected 11 targets for follow-up Carbon Monoxide (CO) and far-infrared (FIR) observations, indicated as stars in Fig.~\ref{fig:pc12}:

\begin{itemize}

\item We targeted 4 starburst ages (${\rm t_{SB}}$), either side of the observed discontinuities in L(H$\alpha$) (see Section \ref{sec:opt}): 3 galaxies with ${\rm t_{SB}}<50$\,Myr, 2 with ${\rm
    t_{SB}}\sim100$\,Myr, 2 with ${\rm t_{SB}}\sim400$\,Myr, and 3
  with ${\rm t_{SB}}\sim600$\,Myr. To extend the time baseline probed
  by our observations, we additionally targeted an older
  PSB galaxy, which is too old to form part of the complete
  sample described above (open star in Fig. \ref{fig:pc12}). As discussed above, typical errors on ${\rm t_{SB}}$ are significantly less than the difference in ages between the bins.
  
\item All objects have typical \halpha\, luminosities for their starburst
  age (see Section \ref{sec:opt}).
  
\item We ensured that galaxies in each of the 4 samples have similar
  stellar mass\footnote{During selection, the stellar mass was measured using a Bayesian fit of stellar population synthesis models to the five-band SDSS photometry (J. Brinchmann; http://www.mpa-garching.mpg.de/SDSS).} ($9.6<{\rm log_{10}(\mstar/M_\odot)}<10.4$) and redshift ($0.025<z<0.05$). 
  
\end{itemize}

Table \ref{tab:obsns} gives the name, position, SDSS spectroscopic ID,
and redshift of all targets. The final column gives the integration
time for the CO observations (Section \ref{sec:co}). 

The $r$-band half-light radius (Petrosian $R_{50}$) of the sample ranges from 1.5--5.2\arcsec\ (1.0--3.4\,kpc in physical radius).
It is important to note that our sample has been selected primarily on
properties derived from spectra which probe the central 3\arcsec\ of
the galaxies (0.9--1.4\,kpc in physical radius), which extends from 0.3--1.0 times the half-light radius.
We take care to note this where relevant to our results.

\begin{figure*}
\includegraphics[scale=0.7]{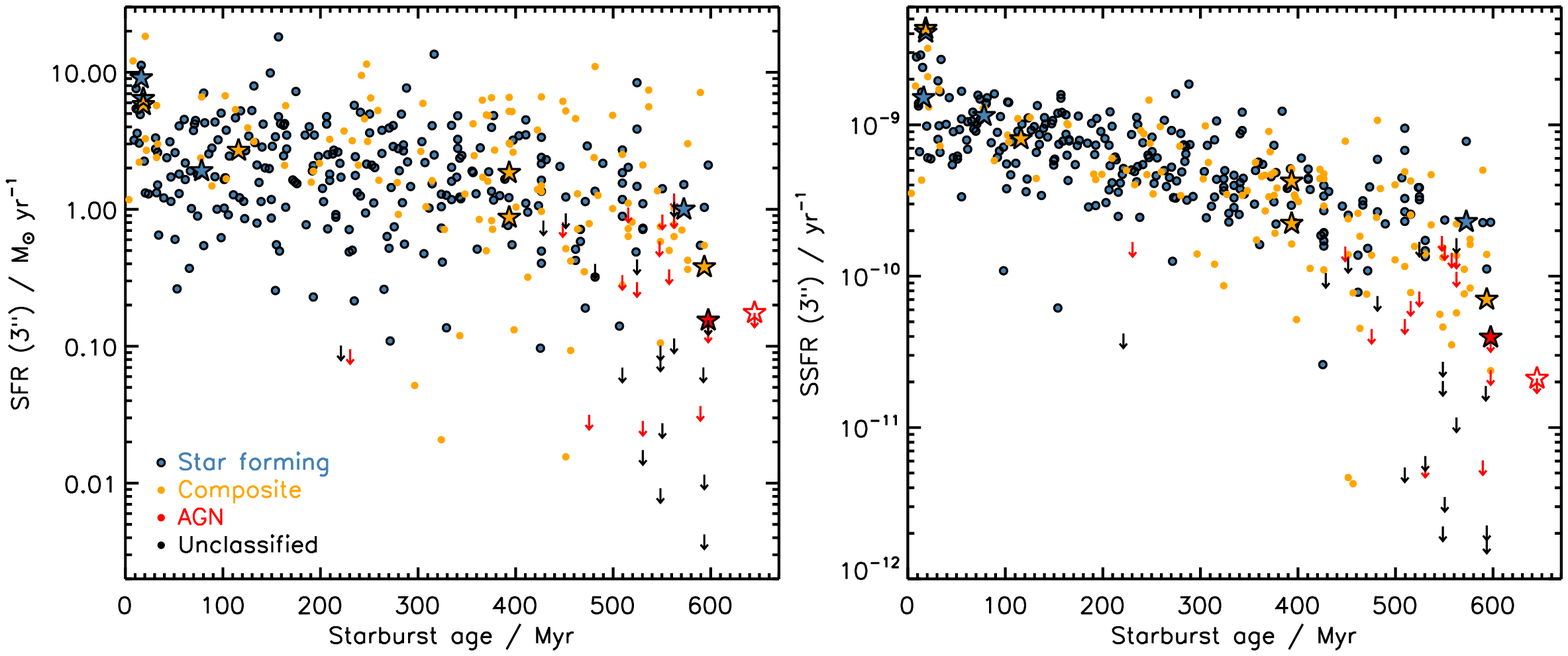}
\caption{The SFR (SFR, left panel) and specific SFR
  (SSFR = SFR/\mstar, right panel) within the 3\arcsec\ SDSS fibre of the
  targeted galaxies (stars) and full starburst sample (dots), as a
  function of time since the starburst. Galaxies are classified as
  star-forming (blue), composites (orange) or AGN (red) on the BPT diagram \citep{BPT81, VO87}. Composite galaxies have had the AGN
  contribution to their \halpha\, luminosity removed, based on the
  correction presented in WHC10, before their SFR is calculated. A
  similar correction is not attempted for the AGN, and these are shown
  as upper limits based on their total \halpha\, luminosity. `PSB11' has a
  starburst age of $\sim$1Gyr, and is indicated as an open star at the right-most edge
  of each panel.}
  \label{fig:ssfr}
\end{figure*}

Fig. \ref{fig:ssfr} shows how the central SFR (measured from dust-corrected \halpha\, luminosity within the 3'' SDSS
fibre) declines with starburst
age as derived from the stellar continuum.  Overall, we measure a decline
timescale of $\sim300$\,Myr similar to that expected from disc
dynamical timescales \citep{LehnertHeckman96}. However, we clearly
identify 3 phases: a short ``starburst'' phase of order 10--30\,Myr, an
almost flat ``coasting'' phase which lasts to around 400\,Myr, and a
subsequent decline. It is clear, even from the SDSS data, that star formation has not shut down instantaneously following the starburst, and residual star formation continues long into the PSB phase. The aim of the CO and FIR observations presented in this paper is to study the galaxy-wide reservoir of molecular gas that is fuelling this ongoing star-formation.

\section{Data and methods}
\label{sec:data}

\subsection{Optical data}
\label{sec:opt}

\begin{table*}
  \centering
  \caption{\label{tab:opt} Physical parameters derived from SDSS data. Note that starburst age, stellar mass, SFR, metallicity and emission line class are measured within the central 3\arcsec\ (0.9-1.4\,kpc in physical radius). Composite galaxies have had the AGN contribution to their \halpha\, luminosity removed, based on the
  correction presented in WHC10, before their SFR is calculated. A
  similar correction is not attempted for the AGN, for which SFR is given as an upper limit based on their total \halpha\, luminosity.}

  \vspace{0.1cm}
  \begin{tabular}{ccccccc}\hline
   Name & $\sigma^*$ & Starburst Age  & log$_{10}$(Stellar Mass) &
   12+log(O/H)$^a$  & SFR & Emission line class  \\ 
    & (km s$^{-1}$) & (Myr) & (M$_\odot$) & & (M$_\odot$ yr$^{-1}$)  & \\ \hline 

    PSB1 &           86$\pm$          13& $          18^{+           4}_{-
           8}$ &9.8&8.9&
$6.4^{+0.4}_{-0.4}$ & SF \\
PSB2 &          131$\pm$          10& $          16^{+           2}_{-
           6}$ &10.7&9.0&
$9.1^{+0.4}_{-0.5}$ & SF \\
PSB3 &          114$\pm$          10& $          18^{+           1}_{-
           4}$ &9.6&9.0&
$\sim5.8$ & comp \\
PSB4 &           96$\pm$          12& $          78^{+          15}_{-
          31}$ &10.4&8.8&
$1.9^{+0.1}_{-0.2}$ & SF \\
PSB5 &           88$\pm$           8& $         116^{+          36}_{-
          65}$ &9.9&-99.9&
$\sim2.7$ & comp \\
PSB6 &          110$\pm$           7& $         394^{+          38}_{-
          74}$ &10.4&-99.9&
$\sim1.8$ & comp \\
PSB7 &           76$\pm$           7& $         394^{+          36}_{-
          63}$ &10.2&9.1&
$\sim0.9$ & comp \\
PSB8 &           96$\pm$           6& $         598^{+         111}_{-
          82}$ &10.0&-99.9&
$<0.2 $ & AGN \\
PSB9 &           87$\pm$           8& $         594^{+         514}_{-
          12}$ &10.3&9.0&
$\sim0.4$ & comp \\
PSB10 &           93$\pm$          15& $         573^{+         165}_{-
          57}$ &10.2&8.9&
$1.0^{+0.1}_{-0.1}$ & SF \\
PSB11 &          101$\pm$           6& $         996^{+          55}_{-
         169}$ &10.5&-99.9&
$<0.2 $ & AGN \\

    \hline
  \end{tabular}
    \begin{minipage}{\textwidth}
      $a$: Gas phase oxygen abundance from \citet{Tremonti04}, -99.9 where unmeasured for
      galaxies with AGN signatures. 
      Note that due to a slightly less
      stringent demarcation between star-forming and composite
      galaxies used in \citet{Tremonti04}, some galaxies we
      class as ``composite'' are classed as star-forming in
      \citet{Tremonti04} and therefore have a metallicity
      measurement.
  \end{minipage}
\end{table*}

The Sloan Digital Sky Survey (SDSS) is an optical photometric and spectroscopic survey of local
galaxies \citep{SDSS_York2000}. The final data release of SDSS-II \citep[DR7,][]{SDSS_DR7}
includes 9380 sq. degrees with spectroscopic coverage, targeting
nearly $10^6$ galaxies with Petrosian $r$-band magnitudes
$<$17.77. The spectra have good signal-to-noise and moderate 
resolution which allows the deconvolution of stellar continuum from
nebular emission that is crucial to accurately measure Balmer
decrements. Such a deconvolution has been carried out by
\citet{Brinchmann04} and \citet{Tremonti04} and the
emission line measurements are available at
http://www.mpa-garching.mpg.de/SDSS.

From the SDSS data, we extract the following information:

\begin{itemize}

\item Model magnitudes in the $u$, $g$, $r$, $i$ and $z$ bands as an input to the {\sc magphys} SED fitting\footnote{To the $u$ and $z$ bands we apply a small correction of -0.04 and +0.02 magnitudes, respectively, 
to bring them onto the AB magnitude calibration (http://classic.sdss.org/dr7/algorithms/fluxcal.html\#sdss2ab).}.
\item Nebular emission line fluxes in the central 3\arcsec, corrected
  for underlying stellar continuum absorption, from \citet{Brinchmann04} and
  \citet{Tremonti04}. These provide the \halpha\, and \hb\ line
  fluxes to estimate dust-corrected \halpha\ luminosity, and allow
  detection of a possible AGN using the BPT diagram
  \citep[\nii/\halpha\, and \oiii/\hb\ emission line ratios;][]{BPT81, VO87}.
  
\item Stellar continuum indices measured in the central 3\arcsec, PC1
  and PC2, which measure the 4000\AA\ break strength and the Balmer
  absorption lines as described in Section \ref{sec:sample} \citep{Wild07}.

\item Stellar surface mass density (\mustar) calculated as $\mstar/(2
  \pi r_{50,z}^2)$ where $r_{50,z}$ is the physical size in kpc of the radius which
  contains 50\% of the $z$-band Petrosian flux. 

\end{itemize}

In Table \ref{tab:opt} we give the parameters derived from SDSS data
that are most relevant to this paper for each object in our
sample. 
The multicolour $gri$ SDSS images of the target galaxies are shown in
Appendix~\ref{fig:cospec} with the size of the 3\arcsec\
SDSS fibre and the Institut de Radioastronomie Millim\'{e}trique (IRAM) beam size at 115 and 230\,GHz. The
images are ordered by starburst age
indicated at the top of each panel. Visual inspection of the images reveals that galaxies
with the youngest starbursts are generally extremely blue, with clear
signs of recent disruption. The host galaxies of older starbursts are
redder, with occasional signs of past disruption (see Pawlik et~al. in prep). For a larger selection of images of the oldest and youngest starbursts in the
parent sample, see Figures A1 and A2 in \citet{Wild10}. 

The SDSS spectra are shown in Figure~\ref{fig:spec}, over the full
optical range in the right-hand column and zoomed in on the 4000\AA\
break region in the left-hand column. The PCA fits to the 4000\AA\
break region, from which the stellar indices PC1 and PC2 and in turn
the starburst ages are measured, are overplotted in red. The emission
line classification of each object is given in the right hand panels,
based upon their position on the BPT diagram. Galaxies that lie below
the demarcation line of \citet{Wild10} are classified as
star-forming, those lying above the demarcation line of
\citet{Kewley01} are classified as AGN, and those lying in
between are classified as star-forming-AGN composites. We note that
these classes are artificial: firstly, the lines are either
empirically derived in the case of \citep{Wild10}, or derived
from theoretical models subsequently shown to be inaccurate
\citep{Levesque10}; and secondly the position of a galaxy in
the diagram depends on the SFR relative to black hole
accretion rate in the 3\arcsec\ SDSS fibre. However, they provide a
useful first order classification between objects where the nebular
emission lines are dominated by star-formation and therefore only a
weak AGN can exist, or those with a rapidly accreting AGN. It should
be noted that pure-AGN are not visible until the older PSB
stage, due to the high SFRs at young ages. See
\citet{Wild10} for more details.

\subsection{CO Data}\label{sec:co}

\begin{table*}
  \centering
  \caption{\label{tab:co} Object name, line intensity, line width,
    velocity offset from systemic (from the optical spectra), gas masses derived from the CO(1-0) line. These are the values measured from the WILMA spectra and do not include a small aperture correction. The final column lists the CO(1-0) aperture correction based on $g$-band isophotal radius (see text for details) which can be applied to correct the gas mass to total.}
  \vspace{0.2cm}
  \begin{tabular}{l D{,}{\pm}{1} D{,}{\pm}{9.0} D{,}{\pm}{9.0} D{,}{\pm}{9.0} D{,}{\pm}{9.0} D{,}{\pm}{9.0} D{,}{\pm}{9.0} D{,}{\pm}{11.0}}
\hline
  \multicolumn{1}{p{0.6cm}}{Name}
& \multicolumn{1}{p{1.4cm}}{\hfil $I_{1-0}^a$}
& \multicolumn{1}{p{1.2cm}}{ $\Delta V_{(1-0)}$}
& \multicolumn{1}{p{1.2cm}}{ $\delta v_{sys (1-0)}$}
& \multicolumn{1}{p{1.4cm}}{\hfil $I_{2-1}^b$}
& \multicolumn{1}{p{1.2cm}}{ $\Delta V_{(2-1)}$}
& \multicolumn{1}{p{1.2cm}}{ $\delta v_{sys (2-1)}$}
& \multicolumn{1}{p{1.0cm}}{\hfil \hfil M$_{\rm gas}$} 
& \multicolumn{1}{p{1.0cm}}{\hfil \hfil \hfil \hfil $f$} \\
  \multicolumn{1}{p{0.6cm}}{}
& \multicolumn{1}{p{1.4cm}}{\hfil (Jy km s$^{-1}$)}
& \multicolumn{1}{p{1.2cm}}{ (km s$^{-1}$)}
& \multicolumn{1}{p{1.2cm}}{ (km s$^{-1}$)}
& \multicolumn{1}{p{1.4cm}}{\hfil (Jy km s$^{-1}$)}
& \multicolumn{1}{p{1.2cm}}{ (km s$^{-1}$)}
& \multicolumn{1}{p{1.2cm}}{ (km s$^{-1}$)}
& \multicolumn{1}{p{1.0cm}}{\hfil \hfil $10^9$M$_\odot$} 
& \multicolumn{1}{p{1.0cm}}{} \\
\hline    
    PSB1 & <1.4 & -- & -- & 13.2\pm2.0 & 87.4\pm11.8 & -5.1\pm11.8 & <0.3 & 
1.04\pm0.28\\
PSB2 & 13.8\pm1.0 & 146.9\pm8.9 & -13.8\pm8.9 & 19.0\pm3.1 & 125.0\pm16.4 & 
83.1\pm16.3 & 3.6\pm0.3 & 1.12\pm0.50\\
PSB3 & 1.4\pm0.4 & 127.4\pm0.0 & 0.0\pm0.0 & 7.8\pm2.0 & 127.4\pm0.0 & 0.0\pm0.0
 & 0.3\pm0.1 & 1.08\pm0.39\\
PSB4 & 12.6\pm1.4 & 88.9\pm9.1 & 22.3\pm9.1 & 35.2\pm2.6 & 91.5\pm5.9 & 
31.5\pm5.9 & 2.1\pm0.2 & 1.12\pm0.49\\
PSB5 & 11.5\pm0.6 & 98.6\pm4.8 & 3.1\pm4.8 & 27.1\pm2.9 & 102.1\pm9.7 & 
52.1\pm9.7 & 2.6\pm0.1 & 1.03\pm0.22\\
PSB6 & 11.3\pm1.3 & 107.5\pm10.8 & 14.3\pm10.8 & 27.7\pm3.6 & 81.1\pm9.5 & 
18.3\pm9.5 & 4.1\pm0.5 & 1.05\pm0.31\\
PSB7 & 6.8\pm1.1 & 122.1\pm16.4 & -0.4\pm16.4 & 7.6\pm1.7 & 66.3\pm13.0 & 
1.2\pm13.0 & 3.0\pm0.5 & 1.02\pm0.19\\
PSB8 & 1.8\pm0.4 & 65.0\pm10.5 & -164.7\pm10.5 & <5.3 & -- & -- & 0.8\pm0.2 & 
1.02\pm0.18\\
PSB9 & 6.6\pm0.7 & 129.5\pm12.2 & 36.1\pm12.2 & <7.3 & -- & -- & 3.1\pm0.3 & 
1.04\pm0.28\\
PSB10 & 8.0\pm0.9 & 132.1\pm12.9 & -1.7\pm12.9 & 12.9\pm1.6 & 64.6\pm7.2 & 
77.4\pm7.2 & 1.9\pm0.2 & 1.08\pm0.39\\
PSB11 & 1.4\pm0.3 & <42.5 & -71.7\pm7.9 & 6.3\pm1.4 & 127.4\pm0.0 & 
0.0\pm0.0 & 0.7\pm0.1 & 1.07\pm0.36\\

    \hline
  \end{tabular}
\end{table*}

We observed all objects on the nights of 2010 June 29/30 with the IRAM-30m
telescope. The Eight MIxer ReceiveR (EMIR) was used in wobbler switching mode with reference positions offset by $\pm100$\arcsec. We used the Wideband Line Multiple
Autocorrelator (WILMA) backend with the standard instrument setup which allowed
us to obtain simultaneous measurements of CO(1-0) at 3mm and CO(2-1) at 1mm,
with a resolution of 2 MHz ($\sim 5$\,kms$^{-1}$ at the observing frequency). We
also simultaneously recorded data at 3mm with the VErsatile SPectrometer  Assembly (VESPA) and with the 4MHz filterbank
at 1mm as a backup.
In general, measurements from the different receivers are
consistent within the $1\sigma$ errors. Conditions were generally good, with
only 2 hours lost due to rain or excessive cloud. The precipitable water vapour at 3 mm ranged from $4.7-37.8$\,mm, with a mean of 13.9\,mm. The total on-source
integration time in minutes is shown in Table \ref{tab:obsns}.

Scans of 6 minutes duration were performed, and were comprised of 12 individual on-source and off-source sub-scans of 30 seconds duration. Observations were interspersed with pointing calibrations
taken every 1--3 hours on Mars or a bright nearby quasar. Telescope refocusing was carried out approximately every
6 hours and following sunset and sunrise. We used the standard reduction
software, {\sc MIRA}, to calibrate each science scan using the calibration scan
taken closest in time. Integrations were repeated until a signal was detected,
or a sufficient upper limit on a line detection was reached. Multiple
integrations were combined in the software package, {\sc CLASS}, using sigma
weighting following first order baseline subtraction. In general baselines were
calculated using a linear fit in line-free regions typically between [-500,-200] and [200,500]\,km s$^{-1}$,
although in a few cases we determined that higher order polynomials were
required following visual inspection of the spectra. A small number of scans on
PSB1 with significantly higher noise levels were discarded, possibly caused by
the worse conditions in which they were taken.

The averaged spectra were smoothed and the channels binned to a velocity resolution of 21\,km s$^{-1}$. Line intensities were measured by summing over the line profile, with a velocity width defined by the FWHM of a
Gaussian fit to the line profile (using least squares minimisation with width, amplitude and
centroid free to vary). We estimate the line intensity using
a sum under the line profile. 
Following \citet{Young11} and \citet{Saintonge11a} in the case of a non-detection/weak line (i.e. $<4\sigma$) we calculated upper limits by summing over a velocity width of 300 km s$^{-1}$ centred on the systemic velocity of the galaxy measured from the SDSS spectrum. 

We determined the statistical uncertainty on each line intensity ($\sigma_l^2$) following \citet{SageWelchYoung07} and \citet{Young11} as 
\begin{equation}
\sigma_l^2 = (\Delta v)^2 \sigma^2 N_l \left( 1 + \frac{N_l}{N_b} \right) ,
\end{equation}

\noindent for a line of width $\Delta v$ with $N_l$ line channels and $N_b$ baseline channels.
The rms noise level $\sigma$ is equal to the standard deviation of the flux measured in line-free regions. Where a line was not detected at $>4\sigma$ we define an upper limit as four times 
the statistical uncertainty of the integrated line intensity. 
10/11 PSBs were detected with $>4\sigma$ significance in the CO(1-0) transition and 9/11 PSBs in the CO(2-1) transition.
Table \ref{tab:co} presents the CO line measurements and Fig. \ref{fig:cospec} shows the observed spectra.

Conversion from antenna temperature to main beam brightness temperature was
achieved using the ratio of the beam and forward efficiencies $B_{\rm{eff}}/F_{\rm{eff}} = 0.83$ at 2.6 mm and 0.64 at 1.3 mm\footnote{http://www.iram.es/IRAMES/mainWiki/Iram30mEfficiencies.}. We then converted from main beam
temperature to flux assuming a factor of 4.73\,Jy\,K$^{-1}$. Gas masses were estimated from the CO(1-0) line flux using a Galactic CO to H$_2$ conversion factor \alphaco$ = 3.2\,\rm{M_{\odot}\, (K\,km\,s^{-1}\,pc^{2})^{-1}}$, with an additional factor of 1.36 for helium. 

Given the optical extent of the galaxies, it is possible that the CO beam ($\sim22$\arcsec at 115\,GHz and $\sim11$\arcsec at 230\,GHz) does not encompass all of the emission. We compute
aperture corrections ($f$) following \citet{Lisenfeld11, Stark13} assuming that the CO distribution follows an exponential disc with scale length given by the $g$-band isophotal radius. The aperture corrections are typically small, with a median correction of 10\% at 115\,GHz and 20\% at 230\,GHz. Where we aperture correct CO fluxes and gas masses we add in quadrature to the errors 5\% of the $g$-band size, to account for the uncertainty of aperture correction.
Whether we choose to apply aperture corrections or not does not change
our conclusions. This is because the optical size of the PSBs is often comparable to or slightly smaller than the half-power beam width at 115\,GHz, meaning that the aperture correction is small. In the following results we use aperture corrected line fluxes
and gas masses unless explicitly stated.

\subsection{Herschel Data}\label{sec:Herschel}

The \emph{Herschel} \citep{Pilbratt10} data were obtained from the OT1\_vwild\_1 open time programme.
All PSBs were observed with the PACS \citep{PACS10} and SPIRE
\citep{Griffin10} instruments in scan map mode in 6 bands at 70, 100, 160, 250, 350 and $500\mu$m with a medium scan speed of 20 arcsec s$^{-1}$. We use the {\sc scanamorphos} map-maker \citep{Roussel13}, implemented in version 12 of HIPE \citep{Ott10}, to mitigate the low frequency 1/$f$ noise in the PACS data, whilst preserving extended source emission. The same version of HIPE was used to create the SPIRE maps, with the na\"ive map-maker being used in this case.

PACS and SPIRE fluxes were determined by summing the flux within a circular
aperture centred on the peak flux of the source. 
We used apertures with radii of 12", 12", 22", 22", 30" and 40" corresponding to
12, 12, 11, 3.7, 3.0 and 2.9 pixels at 70, 100, 160, 250, 350 and $500\mu$m,
respectively. To estimate the background we used annuli with inner and outer radii of 35" and 45" for  flux densities at $70-160\mu$m and 60" and 90" for flux densities at $250-500\mu$m. We applied the aperture corrections as recommended in the PACS\footnote{http://herschel.esac.esa.int/Docs/PACS/html/pacs\_om.html.} and SPIRE\footnote{http://herschel.esac.esa.int/Docs/SPIRE/html/spire\_handbook.html.}
observers manuals for the FM7 calibration \footnote{1.247, 1.289, 1.224, 1.261, 1.226, 1.202.}.
For SPIRE sources we assume beam areas of
436, 772 and 1590 arcsec$^2$ at 250, 350 and $500\mu$m, which include a correction
for point sources with a flux density assumed to be dependant on the spectral index as $S_{\nu} \sim \nu^{3.5}$. 
One object, PSB8, was
found to lie in a crowded field, with the standard aperture at $160\mu$m
overlapping a neighbouring source. For this object we reduced the 
aperture radius by a factor of two for the $160\mu$m observations, increasing the
aperture correction accordingly. For observations at $\geq 250\mu$m we set the fluxes to upper limits at the measured flux values.
The observing strategy required to obtain  both 70 and 100$\mu$m maps means that 
two independent maps were obtained at 160$\mu$m. We treated the two $160\mu$m maps
independently, deriving flux densities which were combined using
sigma weighting. In all cases the independent $160\mu$m fluxes agree to within
the $1\sigma$ errors. We also apply the standard colour corrections for point sources from the SPIRE observers manual of 0.8930, 0.8978, 0.8757
assuming that the flux density can be described as $S_{\nu} \sim \nu^{3.5}$ in the Rayleigh-Jeans (R-J) regime.

We determined the uncertainties on the PACS flux densities following \citet{Balog14}, by measuring the signal rms in the background annulus used for sky subtraction in each band. We correct for the correlated noise between pixels which results from regridding the data from the detector pixel size onto pixels projected on the sky. The noise increases by a factor of 3.4, 3.2 and 8.0. We use this method of error estimation as it results in a higher (more conservative) estimate of the noise than those using the standard deviation of the flux density measured in 6 sky apertures.

For the SPIRE fluxes, following \citet{Ciesla12} the total error ($\sigma_{\rm{tot}}$) encompasses the instrumental error
($\sigma_{\rm{inst}}$), the sky background error ($\sigma_{\rm{sky}}$) and the confusion error ($\sigma_{\rm{conf}}$), and is calculated as:

\begin{equation} 
\sigma_{\rm{tot}} = \sqrt{ \sigma_{\rm{inst}}^2  + \sigma_{\rm{sky}}^2 +
						\sigma_{\rm{conf}}^2}.
\end{equation}

\noindent For SPIRE data $\sigma_{\rm{inst}}$ was calculated by summing in quadrature the values on the error map within the same aperture as the source. 
For each SPIRE band we estimated the sky background noise level ($N_{\rm{pix}} \sigma_{\rm{skymean}}$) by measuring the
standard deviation of the mean sky value in each aperture around the source, using 16 apertures at 250 and $350\mu$m, and 8 apertures at $500\mu$m with the same size as the source aperture, where $N_{\rm{pix}}$ is the number of pixels in the aperture.
For the SPIRE maps we adopt the confusion noise estimates ($err_{\rm{conf}}$) from \citet{Nguyen10} of 5.8, 6.3 and 6.8 mJy/beam at 250, 350 and $500\mu$m, respectively. 
The confusion error is

\begin{equation} 
\sigma_{\rm{conf}} = err_{\rm{conf}} \times \sqrt{ \frac{N_{\rm{pix}} \times pixsize^2}{beamarea} },
\end{equation}

\noindent following \citet{Ciesla12}. The confusion noise at the depth of the PACs maps is negligible. We convolve the uncertainties in quadrature with a flux calibration error of 5\% and 5.5\% of the PACS\footnote{Note that due to the uncertainty in the appropriate colour correction, the PACS flux could be uncertain by up to 20\%.} and SPIRE fluxes, respectively, as recommended in the PACS and SPIRE manuals, \citep{Bendo13}. 
Following \citet{Smith13} we use all positive flux measurements in each
band regardless of the signal-to-noise in order to include as much information as possible in the SED. Where a negative flux is measured we include the flux as an upper limit consistent with a flux density of zero.

\subsection{Ancilliary data}\label{sec:ancilliarydata}

For each source we compile $FUV$ and $NUV$ data from the \emph{GALEX} DR7
catalogue server. All sources are matched to the nearest source within a 3"
radius. We reject fluxes which have artifact flags and where the source is 
too close to the detector edge for reliable flux measurements (field-of-view radius $>0.55\deg$). 
Where multiple observations exist we select the deepest observation. We
correct the UV fluxes for galactic extinction using the values of $E(B-V)$ from
\citet{SFD98} and the factor $A_{FUV}=7.9E(B-V)$ and
$A_{NUV}=8.0E(B-V)$, derived using the \citet{Cardelli89} Galactic extinction law for a total-to-selective extinction ratio of $R_V=3.1$. Non-detections
($<5\sigma$) are included as upper limits in the SED fitting.

We compile near-infrared $J$, $H$, and $K$ elliptical isophotal
magnitudes from the Two Micron All Sky Survey \citep[2MASS;][]{Skrutskie06}
Extended Source Catalog \citep[XSC;][]{Jarrett00}.
We also compile \emph{Wide-field Infrared Space Explorer} \citep[\emph{WISE;}][]{Wright10} $3-22\mu$m fluxes from the all-sky data release. We use the profile mags (\emph{w4mpro}) for point sources
(defined with ${\rm{ext\_flg}}=0$), and the aperture magnitudes (\emph{w4gmag}) for
extended sources. Following \citet{Jarrett12} we add small zeropoint corrections of 0.03, 0.04, 0.03, and -$0.03$ to the WISE W1,
W2, W3, and W4 magnitudes, respectively. We convert the WISE magnitudes to fluxes including a colour correction dependant on the spectral slope
\footnote{http://wise2.ipac.caltech.edu/docs/release/allsky/expsup/sec4\_4h.html.}.
We apply a further colour correction to the W4 band as our sources have rising spectral slopes in this band, following the updated calibration in \citet{Brown14}.
Where fluxes are not detected in the WISE bands we utilise upper limits in the SED fitting at the $2\sigma$ level.

We compiled fluxes from the \emph{IRAS} Faint Source Catalogue \citep{FSC89} at 25 and 60$\mu$m which have a quality flag $>2$. Where fluxes were of low quality or undetected we used $5\sigma$ upper limits of 0.2\,Jy in the SED fitting.

In the UV, optical, near- and mid-infrared fluxes we convolve the catalogue error in quadrature with a calibration error of 20, 10, 15 and 10\% of the flux respectively, to allow for differences in the methods used to measure total photometry and errors in the spectral
synthesis models used to fit the underlying stellar populations. 
To account for larger calibration uncertainty in the WISE W4 band we convolve the 
catalogue error in quadrature with 20\% of the measured, colour-corrected flux.

\subsection{SED fitting}
\label{sec:SEDfitting}

The wealth of multi-wavelength data for our sample of PSBs allows
us to derive global physical properties by fitting models to their broad-band photometric SEDs. We use the physically motivated method of \citet*[][hereafter
DCE08\footnote{The \citet*{DCE08} models are publicly available as a
user-friendly model package {\sc magphys} at www.iap.fr/magphys/.}]{DCE08} to
recover the physical properties of the PSBs. 

DCE08 employ an energy balance prescription, whereby the UV--optical radiation emitted by stellar populations is absorbed by dust, and this absorbed energy is matched to that re-radiated in the FIR. The optical library of 50,000 spectra is produced using the the latest version of the population synthesis code of \citet{BC03}, Charlot \& Bruzual (2007, in prep), and assumes exponentially declining SFHs with additional superimposed random bursts (known as stochastic SFHs). The model spectra cover a wide range of age, metallicity, SFH and dust attenuation and a \citet{Chabrier03} is assumed. The infrared libraries contain 50,000 SEDs comprised of four different temperature dust components. In
stellar birth clouds, these components are polycyclic aromatic hydrocarbons
(PAHs), hot dust (stochastically heated small grains with a temperature
$130-250$\,K), and warm dust in thermal equilibrium ($30-60$\,K). In the diffuse
ISM the relative fractions of these three dust components are fixed, but an
additional cold dust component with an adjustable temperature between 15 and
25\,K is added. The dust mass absorption coefficient $\kappa_{\lambda} \propto
\lambda^{-\beta}$ has a normalisation of $\kappa_{850}=0.077\,\rm{m}^2
\,\rm{kg}^{-1}$ \citep{SLUGS00a}. A dust emissivity index of $\beta=1.5$ is
assumed for warm dust, and $\beta=2.0$ for cold dust.

Statistical constraints on the various parameters of the model are derived using the Bayesian approach described in DCE08, which ensures that possible degeneracies between model parameters are included in the final probability density function (PDF) of each parameter. The effects of individual wavebands on the derived parameters are explored in DCE08 and \citet{DJBSmith12}. The {\sc magphys} code used in this paper has been modified from the public version to take into account flux density upper limits in the $\chi^2$ calculation to give additional constraints on physical parameters, as described in \citet{DJBSmith12} and \citet{Rowlands14a}. For more details of the method we refer the reader to DCE08.

\begin{table*}
  \centering
  \caption{\label{tab:MAGPHYS} Physical parameters derived from fitting the UV to far-infrared SEDs of each galaxy. The columns are (from left to right): Name, $\mstar/\rm{M}_\odot$, log$_{10}$(stellar mass); ${M}_\mathrm{d}/\rm{M}_\odot$,  log$_{10}$(dust mass); \mdms,  log$_{10}$(dust-to-stellar mass ratio); \tbgscold/K, temperature of the cold diffuse ISM dust component; \ldust/$\rm{L}_\odot$,  log$_{10}$(dust luminosity); SFR/$\rm{M}_\odot yr^{-1}$, the log$_{10}$(SFR) averaged over the last $10^8$ years and SSFR/yr$^{-1}$, the log$_{10}$(SSFR) averaged over the last $10^8$ years. Uncertainties are indicated by the median 84th--16th percentile range from each individual parameter PDF. Note that \tbgscold should be treated with caution as the median likelihood values are close to the bounds on the temperature prior.}
  \vspace{0.3cm}
  
  \begin{tabular}{cccccccc}\hline
   Name & \mstar & \md  & \mdms & \tbgscold & \ldust & SFR & SSFR  \\ 
        & log$_{10}$(M$_\odot$) & log$_{10}$(M$_\odot$) & log$_{10}$(\mdms) & (K) & log$_{10}$(L$_\odot$) & log$_{10}$(M$_\odot$yr$^{-1}$) & log$_{10}$(yr$^{-1}$)  \\
   \hline 
    PSB1 & $9.33^{+0.15}_{-0.03}$ &$7.32^{+0.09}_{-0.03}$  &$-2.01^{+0.06}_{-0.16}
$  &$23.8^{+1.2}_{-0.7}$  &$11.00^{+0.02}_{-0.01}$  &$1.04^{+0.04}_{-0.03}$  &$
-8.27^{+0.01}_{-0.15}$ \\
PSB2 & $10.48^{+0.06}_{-0.01}$ &$7.61^{+0.01}_{-0.01}$  &$-2.88^{+0.01}_{-0.05}
$  &$24.6^{+0.1}_{-0.1}$  &$11.36^{+0.01}_{-0.01}$  &$1.21^{+0.15}_{-0.01}$  &$
-9.27^{+0.10}_{-0.01}$ \\
PSB3 & $9.50^{+0.05}_{-0.04}$ &$6.93^{+0.12}_{-0.06}$  &$-2.55^{+0.13}_{-0.10}
$  &$23.8^{+0.7}_{-2.2}$  &$10.62^{+0.04}_{-0.09}$  &$0.74^{+0.05}_{-0.04}$  &$
-8.77^{+0.10}_{-0.05}$ \\
PSB4 & $9.81^{+0.01}_{-0.07}$ &$7.45^{+0.03}_{-0.01}$  &$-2.34^{+0.07}_{-0.03}
$  &$24.3^{+0.5}_{-0.4}$  &$10.84^{+0.01}_{-0.02}$  &$0.74^{+0.01}_{-0.02}$  &$
-9.07^{+0.10}_{-0.01}$ \\
PSB5 & $9.46^{+0.08}_{-0.06}$ &$7.23^{+0.01}_{-0.01}$  &$-2.24^{+0.08}_{-0.09}
$  &$24.1^{+0.6}_{-0.1}$  &$10.86^{+0.01}_{-0.03}$  &$0.80^{+0.03}_{-0.04}$  &$
-8.68^{+0.10}_{-0.20}$ \\
PSB6 & $10.47^{+0.02}_{-0.15}$ &$7.50^{+0.08}_{-0.14}$  &$-2.95^{+0.11}_{-0.15}
$  &$21.4^{+1.5}_{-0.8}$  &$10.58^{+0.04}_{-0.02}$  &$0.40^{+0.07}_{-0.03}$  &$
-10.07^{+0.24}_{-0.01}$ \\
PSB7 & $10.08^{+0.09}_{-0.02}$ &$6.91^{+0.06}_{-0.05}$  &$-3.18^{+0.08}_{-0.09}
$  &$24.6^{+0.3}_{-0.8}$  &$10.24^{+0.05}_{-0.02}$  &$-0.13^{+0.04}_{-0.02}$  &$
-10.22^{+0.05}_{-0.05}$ \\
PSB8 & $9.89^{+0.15}_{-0.11}$ &$6.41^{+0.27}_{-0.18}$  &$-3.47^{+0.30}_{-0.23}
$  &$22.9^{+1.5}_{-2.5}$  &$9.57^{+0.05}_{-0.08}$  &$-0.94^{+0.18}_{-0.23}$  &$
-10.87^{+0.25}_{-0.25}$ \\
PSB9 & $10.11^{+0.03}_{-0.10}$ &$7.19^{+0.08}_{-0.10}$  &$-2.92^{+0.14}_{-0.12}
$  &$23.1^{+1.1}_{-1.0}$  &$10.23^{+0.03}_{-0.02}$  &$-0.16^{+0.11}_{-0.01}$  &$
-10.22^{+0.05}_{-0.05}$ \\
PSB10 & $10.04^{+0.05}_{-0.23}$ &$7.24^{+0.02}_{-0.02}$  &$-2.79^{+0.22}_{-0.06}
$  &$24.6^{+0.2}_{-0.2}$  &$10.57^{+0.01}_{-0.04}$  &$0.23^{+0.06}_{-0.05}$  &$
-9.82^{+0.30}_{-0.05}$ \\
PSB11 & $10.49^{+0.03}_{-0.06}$ &$7.03^{+0.23}_{-0.27}$  &$-3.46^{+0.23}_{-0.28}
$  &$19.5^{+2.2}_{-1.8}$  &$9.70^{+0.03}_{-0.06}$  &$-0.82^{+0.03}_{-0.07}$  &$
-11.32^{+0.10}_{-0.05}$ \\

    \hline
  \end{tabular}
  
\end{table*}

\begin{figure}
\includegraphics[width=0.48\textwidth]{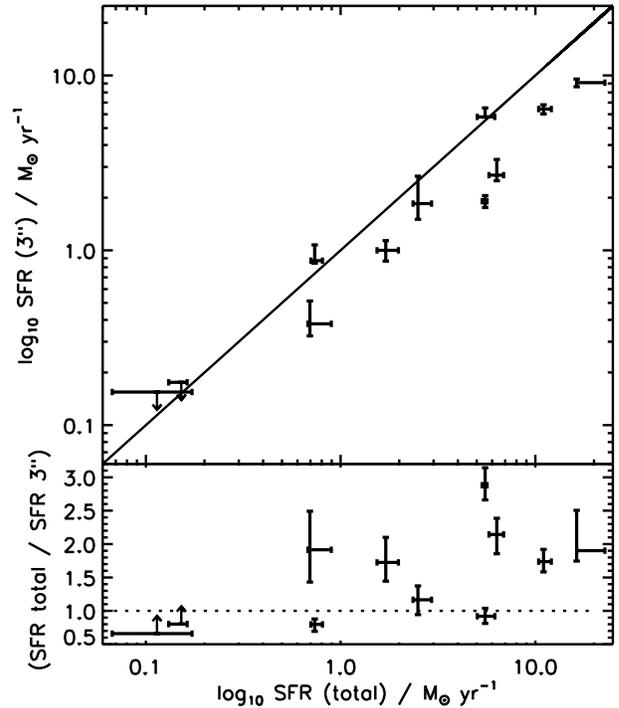}
\caption{Comparison of the total SFR from multi-wavelength SED fitting and the 3\arcsec\ SFR (0.9-1.4\,kpc in radius), from the dust-corrected H$\alpha$ luminosity. The solid line is the one-to-one relation. The lower plot shows the ratio of total SFR to 3" SFR as a function total SFR. In the lower panel we account for the propagation of asymmetric error values via a Monte-Carlo analysis. The total SFR is on average slightly higher than the fibre SFR by around a factor of two.}
\label{fig:compare_sfr}
\end{figure}

The best-fit SEDs are shown in Fig.~\ref{fig:SED_example}, and the fitted values and errors on the main parameters of interest are provided in Table~\ref{tab:MAGPHYS}. In Figure~\ref{fig:compare_sfr} we compare the total SFR derived from the SED fitting (averaged over the last $10^8$ years) to the SFR in the central 3\arcsec\ fibre derived from the dust-corrected H$\alpha$ emission. There is a very good agreement between the two measurements over the wide range of SFR probed, however there is an offset with the total SFR higher than the fibre SFR by around a factor of two on average. This may be due to: (1) the fibre only covering the inner 0.9-1.4\,kpc in radius (30-100\% of the $r$-band half-light radius, and 10-30\% of the $r$-band Petrosian $R_{90}$ radius); (2) the H$\alpha$ derived SFR is not exactly equivalent to a SFR derived from SED fitting, and in particular relates to an average over a shorter time interval; (3) the extinction corrected H$\alpha$ may underestimate the true SFR in the mostly heavily extincted sites of star-formation. It is therefore difficult to draw any firm conclusions about the distribution of star formation in the galaxies from these observations. In the following results we consider the total SFR, unless stated otherwise.

\section{Results}
\label{sec:results}

In WHC10 we used the full sample of local ($z<0.07$) bulge-dominated SDSS galaxies to study the black hole accretion
activity as a function of starburst age (``Timing the starburst-AGN
connection''). Here we wish to address very specifically the question
of \emph{how and why} star formation declines following a starburst, which is currently very poorly constrained observationally.

\subsection{Star-formation efficiency}

\begin{figure}
\includegraphics[width=0.48\textwidth]{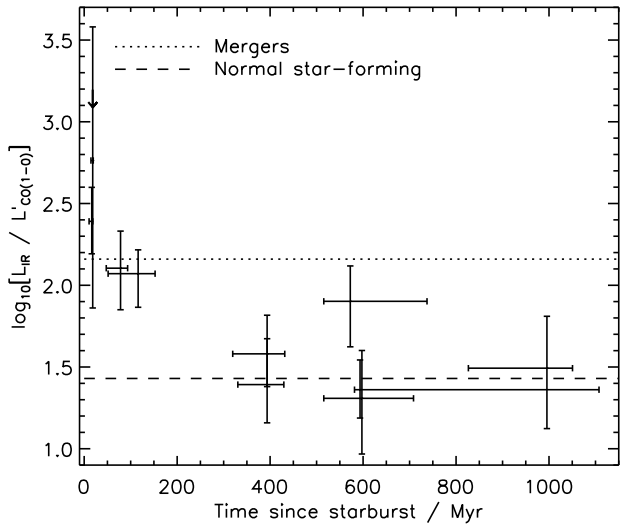}
\caption{The ratio of FIR luminosity (\lfir) to CO(1-0) luminosity (a proxy for SFE), as a function of starburst age. A clear trend of decreasing \lfir/\lco\ with starburst age is observed. The dotted line shows the average value of \lfir/\lco\ for IR luminous mergers and the dashed line shows the average value for normal galaxies from \citet{Genzel10}.}
\label{fig:SFE}
\end{figure}

The star-formation efficiency (SFE) quantifies how quickly gas is consumed by star formation. For consistency with previous work, we begin this section by plotting in Figure~\ref{fig:SFE} the ratio of far-infrared (FIR) luminosity (\lfir; the integrated emission between 3 and $1000\mu$m) to the CO line luminosity (\lco; \citealp{SolomonVandenBout05}), as a function of starburst age. Under the assumption that \lfir\, corresponds to total SFR\footnote{This may not be true for galaxies no longer undergoing a starburst, but this would only reinforce the trend in Fig.~\ref{fig:SFE} since the infrared luminosity would overpredict the SFR (see \citealt{Rowlands14a}).} and \lco\ traces the gas mass, the ratio of these quantities is a proxy for the SFE.

We find that the young PSBs have \lfir/\lco\ values around a factor of 5 higher than the older PSBs. This is similar to the difference found between interacting and isolated galaxies \citep[a factor of 4--6,][]{Young86,Lisenfeld11}. The \lfir/\lco\ of the young PSBs is consistent with IR luminous mergers/starbursts ($L>10^{11}$L$_\odot$) from \citet{Genzel10}, even though the PSBs have a lower FIR luminosity (likely as a result of their lower mean stellar mass compared to typical Luminous Infrared Galaxies (LIRGs; with \mstar$>10^{10}$M$_\odot$). Over a timescale of $\sim 300$\,Myr the \lfir/\lco\ falls, and is consistent with that of normal star-forming galaxies for the older PSBs \citep{Genzel10}. We do not find PSBs with \lfir/\lco\ similar to those of gas-rich ETGs, which have a \lfir/\lco\ a factor of 2.5 lower than normal spiral galaxies \citep{Davis14}.
It is clear that by 0.5--1\,Gyr following the starburst, the star formation is not completely quenched and the star formation remains relatively efficient. It is interesting to note that the time around which the decrease in \lfir/\lco\ occurs is coincident with a significant increase in the black hole accretion rate in PSBs (WHC10), however, this correlation does not imply any causation.

\begin{figure}
\includegraphics[width=0.48\textwidth]{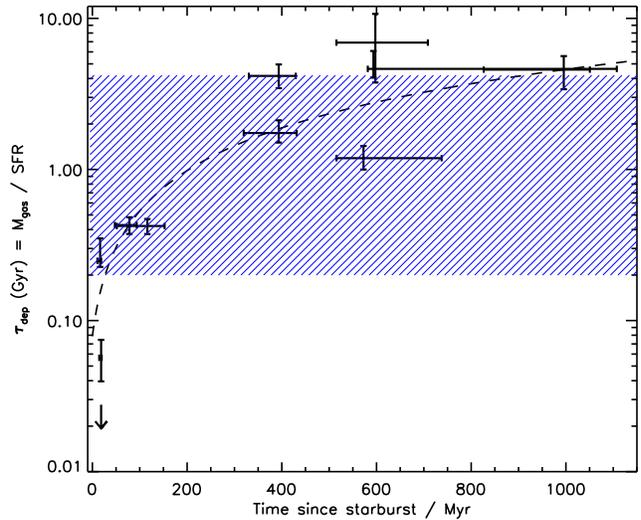}
\caption{The molecular gas depletion time ($\tau_{\rm{dep}} = \rm{M}_{\rm{gas}}/$SFR from {\sc magphys}) as a function of starburst age for the PSBs. The dotted line shows the relation for a linear increase in depletion time with starburst age. The blue hatched area indicates the typical molecular gas depletion time scales for 30 nearby disc galaxies from \citet{Leroy13}.}
\label{fig:Tdep}
\end{figure}

The molecular gas depletion time is inversely proportional to the SFE. In Figure~\ref{fig:Tdep} we plot the molecular gas depletion time in years ($\tau_{\rm{dep}} = \rm{M}_{\rm{gas}}/$SFR) as a function of starburst age. Here we replace \lfir\, with the SFR measured from the full multi-wavelength SED fit, to be consistent with most recent literature. We observe a clear trend of increasing molecular gas depletion time with starburst age, which is best-fit by the linear relation 
\begin{equation}
\tau_{\rm{dep}}/10^9  = 0.0045 \rm{t_{SB}} + 0.078
\end{equation}
where $\tau_{\rm{dep}}$ is in Gyr, and $\rm{t_{SB}}$ is in Myr. We note that this trend is analogous to that seen between \lfir/\lco\ and starburst age in Figure~\ref{fig:SFE}, and therefore the observed trend is independent of the choice of \alphaco. 
We measure depletion times of $\sim1-8$\,Gyrs for the oldest PSBs, in agreement with those of normal spiral galaxies of a similar stellar mass \citep{Leroy08, Bigiel08, Saintonge11b, Leroy13, Huang_Kauffmann14}. This suggests that the depletion times of PSBs are returning to normal after a burst of star formation, although we are unable to tell whether they will continue to decline in the future with the present sample. The younger PSBs have shorter depletion times of 50--500\,Myrs, consistent with those of IR luminous mergers ($\tau_{\rm{dep}} \sim 200$\,Myr; \citealp{Genzel10}).

Many studies have now demonstrated that IR luminous mergers/starburst galaxies exhibit an excess SFR relative to their molecular gas content (and hence have a shorter depletion time), when compared to normal star-forming galaxies \citep{Sanders91, SM96, Gao_Solomon04, Daddi10b, Genzel10}. 
Two alternative scenarios exist to explain this result: 
(1) During the interaction/merger gas may be driven into the centre by tidal torques, leading to higher gas surface densities and more efficient star formation \citep{Braine_Combes93, Downes_Solomon98, Tacconi08}; 
(2) The molecular gas reservoir has been depleted and the starburst quenched, but the FIR emission is still visible \citep{Leroy06}. As our measurement of depletion time does not depend directly on the \lfir, and we have shown that the SFR measured from the multiwavelength SED tracks that measured by dust-corrected \halpha\ luminosity which provides a close to instantaneous measure of SFR, we strongly favour the first option. In principle, the combination of stellar population age measured from the stellar continuum, \halpha\ luminosity and \lfir, provides a powerful constraint on the decay rate of \lfir\, following a starburst. The fact that we find no significant offset in timescales between the decay in star formation efficiency implied by Figures \ref{fig:SFE} and \ref{fig:Tdep} rules out the possibility that the FIR emission remains visible for an extended period following the starburst. However, larger numbers of galaxies would be required to determine precisely the impact of using \lfir\ to infer star formation efficiencies and depletion times.

\subsection{Kennicutt-Schmidt relation}

\begin{figure}
\includegraphics[width=0.48\textwidth]{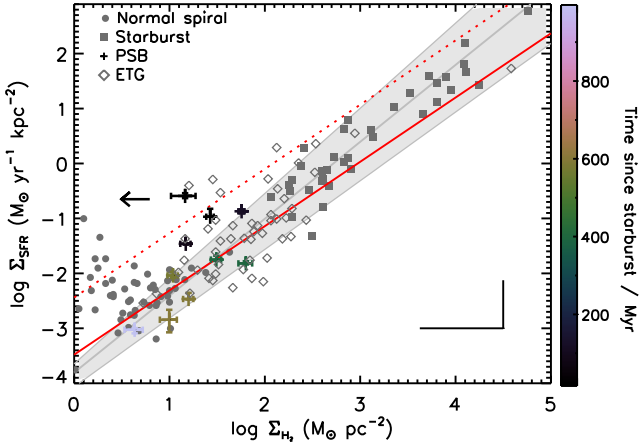}
\caption{The SFR surface density (\sigsfr) as a function of molecular gas mass surface density (\siggas) for the PSBs (crosses), with points colour coded by starburst age. Grey circles show normal star-forming galaxies, and grey squares show circumnuclear starbursts from K98 (corrected down by a factor of 1.6 to a Chabrier IMF). ETGs with molecular gas (and HI gas, although this is negligible)  are shown as unfilled diamonds \citep{Davis14}. All values of \siggas\ assume a Galactic CO conversion factor, and literature values have been corrected to a common value of \alphaco$ = 4.35\,\rm{M_{\odot}\, (K\,km\,s^{-1}\,pc^{2})^{-1}}$. The grey line shows the Kennicutt-Schmidt relation (with the scaling factor corrected downwards by a factor of 1.6 to a Chabrier IMF), with the $1\sigma$ scatter as the grey filled area. The red lines indicate the relation for star-forming galaxies (solid) and mergers (dotted) from \citet{Genzel10}. The black horizontal and vertical bar indicates the median possible systematic in the SFR and gas surface densities when using the $u$-band Petrosian $R_{50}$ instead of the $R_{90}$ radius.}
\label{fig:K-S}
\end{figure}

Gas mass surface density (\siggas) and SFR surface density (\sigsfr) in galaxies are linked by the Kennicutt-Schmidt (K-S) relation \citep{Kennicutt98}.
In Figure~\ref{fig:K-S} we plot the molecular \siggas\ and \sigsfr\ for the PSBs and compare them to normal
star-forming galaxies and circumnuclear starbursts from K98, and ETGs with molecular gas from \citet{Davis14}. For the PSBs we do not have resolved CO or SFR observations so we assume that the CO emitting region is the same as where young stars are located, and use the $u$-band Petrosian $R_{90}$ radius as the area over which we calculate \siggas\ and \sigsfr. This approximation leads to a substantial uncertainty in both parameters: a factor of two decrease in the radius for star-formation and gas emission (i.e going from Petrosian $R_{90}$ to $R_{50}$ radius) shifts the PSBs $\sim1$\,dex to the upper right of the plot.  All \siggas\ estimates assume a Galactic CO-to-H$_2$ conversion factor, although \citet{Genzel10} suggest a lower value of \alphaco\ may be appropriate for IR luminous mergers. Adopting a lower value of \alphaco\ would decrease the gas surface density of the PSBs.

Under the assumptions of size and \alphaco\ made above, we find that the PSBs sit in an intermediate region between normal star-forming spirals and circumnuclear starbursts. There is a clear trend of starburst age relative to position on the K-S relation. The youngest PSBs with an age $<100$\,Myr have the highest SFR and gas mass surface densities, and lie in a region coincident with ETGs with molecular gas \citep{Davis14}, and have slightly lower gas mass and SFR surface densities than local circumnuclear starbursts (K98).
Note that only 22\% of all ETGs in the local Universe ($<42$\,Mpc) harbour a detectable mass of molecular gas, so PSBs have a much higher gas mass surface density than the vast majority of ETGs (those not plotted here). 

The higher gas surface densities of the young PSBs (which are the most morphologically disturbed) relative to normal galaxies is consistent with \citet{Braine_Combes93} and \citet{Downes_Solomon98}, who found higher gas surface densities in interacting systems. The increased gas surface density could be a result of the funnelling of gas into the central regions during interactions/mergers \citep[e.g.][]{Mihos_Hernquist94, Mihos_Hernquist96, Barnes_Hernquist96}.  Observations of dense gas tracers in young PSBs would help confirm this hypothesis \citep{Gao_Solomon04}.
Beyond the $\sim600$\,Myr timespan of our observations, it is unclear whether the PSBs will continue to evolve back to the region occupied by the spirals,
or exhaust their gas supplies to become gas-poor ETGs. More observations of the gas content of PSBs older than 600\,Myr are needed in order to examine the late-time evolution of this population.

Note that we do not consider the total (neutral and atomic) gas surface density in this work, as only 3 PSBs have detections of HI gas (PSB4, PSB5 and PSB11, from \citet{Davoust_Contini04} and the ALFALFA survey (\citealp{Haynes11}), an additional 3 PSBs have upper limits from the Northern HIPASS Catalog \citep{Wong06, Wong09} which do not put strong constraints on the HI gas mass. In normal galaxies there is a large variation in H$_2$/HI mass \citep{Saintonge11b}, therefore we cannot reliably predict HI from H$_2$ in the PSBs. The detection of substantial atomic gas reservoirs in some PSBs indicates that either the HI reservoir has not been depleted \citep{Zwaan13}, or re-accretion of gas can occur after the star formation has been quenched \citep{Saintonge11a, Stark13, YatesKauffmann14}, which may be particularly efficient in low mass systems \citep{Saintonge11b}. Alternatively, HI gas exists in the outer disc and is not affected by the starburst and subsequent quenching event in the central region \citep{Saintonge11b, Pracy14}.

\subsection{Cold ISM content}
\label{ISMContent}

In this section we discuss the physical properties of the cold ISM of PSBs.
From the {\sc magphys} SED fitting we derive a range of dust masses from $0.3-4.0\times{10}^7\,$M$_\odot$, with a median dust mass of $1.7\times{10}^7\,$M$_\odot$. This is similar to that of low redshift spiral galaxies with a similar stellar mass to the PSBs in the \emph{Herschel}-ATLAS \citep{Rowlands12}. The median molecular gas mass derived from the 12CO(1-0) integrated line fluxes is $2.7\times{10}^9\,$M$_\odot$ with a range from $(<0.3-4.3)\times{10}^9\,$M$_\odot$, assuming a Galactic CO to H$_2$ conversion factor. 
The molecular gas masses of the PSBs are similar to those of the most gas-rich ETGs \citep{Young11} and spiral galaxies \citep{Lisenfeld11} in the local Universe (see also \citealt{French15}).

\subsubsection{Gas-to-stellar mass}

\begin{figure}
\includegraphics[width=0.45\textwidth]{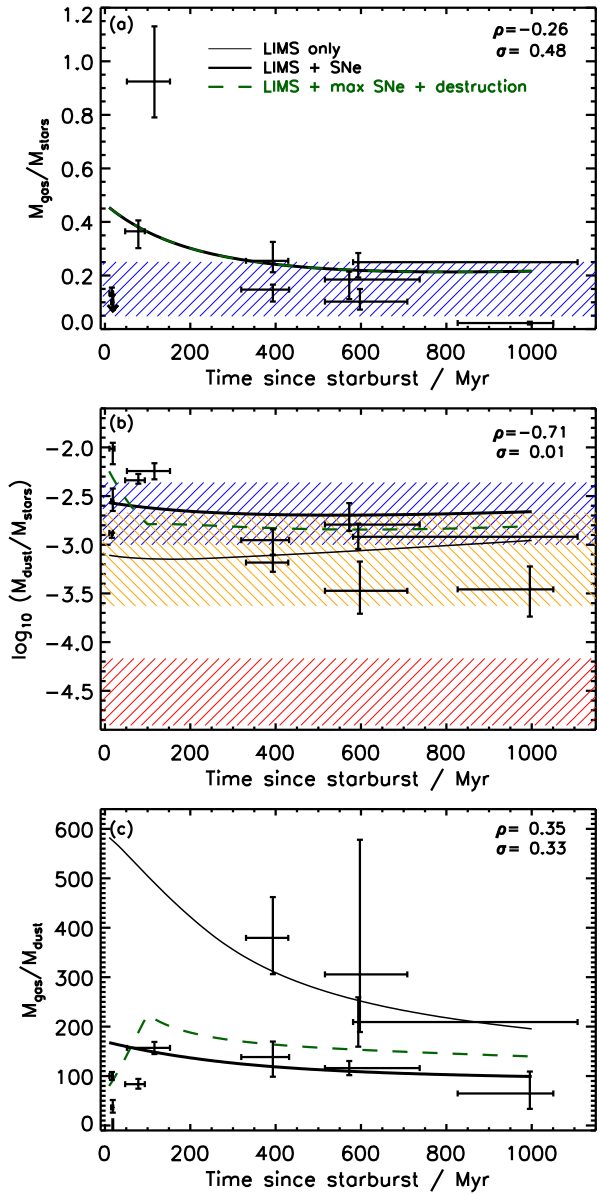}
\caption{(a) Molecular gas-to-stellar mass ratio as a function of starburst age. We observe a wide range of gas-to-stellar mass ratios in both young and old PSBs. The blue hatched region is the range of observed gas-to-stellar mass ratios for normal nearby galaxies ($\sim 0.1$
for CO detected galaxies of a similar stellar mass to the PSBs; $\mstar \sim 10^{10}$M$_{\odot}$; \citealp{Boselli14, Saintonge11a}). 
(b) Dust-to-stellar mass ratio vs starburst age. The blue hatched region is the standard deviation either side of the mean \mdms\, for $z<0.1$ spiral galaxies detected in the \emph{Herschel}-ATLAS \citep{Rowlands12}. The orange hatched region is the standard deviation either side of the mean \mdms\, for $z<0.06$ dusty ETGs from \citet{Agius13}, using updated FIR photometry. The red hatched region is the range of \mdms\, for non-dusty ETGs (representative of red sequence galaxies) for a range of dust temperatures \citep{Rowlands12}. 
(c) Gas-to-dust mass ratio as a function of starburst age. 
Values in the right of each plot indicate the non-parametric Spearman rank coefficient ($\rho$) and the two-sided significance of the deviation of the correlation from zero, or the null hypothesis ($\sigma$, a low value indicates a significant correlation).
On each plot we show closed-box chemical evolution models with dust produced by LIMS only with no destruction (thin black line), LIMS+SNe with no destruction (thick black line), and dust from LIMS+SNe (green dashed line; assuming all metals are converted into dust) with strong destruction from 0-0.1\,Gyr and weak destruction from 0.1-1.0\,Gyr.
Note that the chemical evolution models are not fits to the data.}
\label{fig:ISMcontent}
\end{figure}

In Figure~\ref{fig:ISMcontent} (a) we show the gas-to-stellar mass ratio (M${\rm_{gas}}/$\mstar) of the PSBs as a function of starburst age, which ranges from 0.02 to 0.92 with a median of 0.18. The gas-to-stellar mass ratios of the majority of the PSBs are consistent with the range found for normal nearby galaxies ($\sim 0.1$
for CO detected galaxies\footnote{Note that including CO non-detections decreases the average gas-to-stellar mass ratio to $\sim 0.04$.}; \citealp{Boselli14}) of a similar stellar mass to the PSBs ($\mstar \sim 10^{10}$M$_{\odot}$; see also \citealp{Saintonge11a}). 
The gas-to-stellar mass ratios of the PSBs are on average a factor of 20 higher than those of red sequence galaxies ($NUV-r=6$) which have a typical gas-to-stellar mass ratio of 0.012 \citep{Saintonge11a}. However, local red-sequence samples will be dominated by objects with higher stellar masses, which have lower gas-to-stellar mass ratios. For a detailed comparison, a mass-matched sample would be essential.  
The gas-to-stellar mass ratios of the PSBs are also greater than those of ETGs in the \emph{Herschel} Reference Survey (M${\rm_{gas}}/$\mstar$\sim 5.3 \times 10^{-3}$) \citep{Boselli14} and in the ATLAS$^{\rm 3D}$ sample with detected molecular gas \citep{Cappellari13}. The ATLAS$^{\rm 3D}$ ETGs have gas-to-stellar mass ratios\footnote{Where the stellar mass is the Jeans anisotropic mass, i.e the total stellar mass from dynamical modelling, and should be comparable to our SED derived \mstar\,within a factor of two.} ranging from $8.9 \times 10^{-5}$ to $5.6 \times 10^{-3}$, with an average of $7.1 \times 10^{-4}$. 

Figure~\ref{fig:ISMcontent} (a) also shows that there is no significant correlation of the gas-to-stellar mass ratio with time since the starburst ended (Spearman correlation coefficient $\rho=-0.26$, with the two-sided significance of the deviation of the correlation from zero, or the null hypothesis $\sigma=0.48$), although there is a large scatter in our sample. This suggests that the majority of gas may have already been consumed in the starburst and we are just observing the residual gas reservoir. Alternatively, the global molecular gas reservoir may not be affected by a nuclear starburst event \citep{Pracy14} and only gas in the central region ($\sim$2\,kpc for the galaxies here) may be consumed by star formation or affected by feedback processes from star formation or an AGN. High spatial resolution mapping of the molecular gas is needed to determine this. 

Over time one may expect a slow depletion of the gas reservoir due to star formation, with typical timescales of 0.3\,Gyr related to disc dynamical timescales \citep{Kennicutt98, LehnertHeckman96}. A much larger sample may be required to observe this expected decline, given the typical errors on gas mass measurements at these redshifts. A rapid decline in gas-to-stellar mass ratio at early times, which would indicate expulsion of gas by SNe and/or fast stellar winds, is not observed. A later rapid decline, which would be the smoking gun for the heating or expulsion of gas from an accreting supermassive black hole, is also not observed. Our results disagree with those of \citet{Schawinski09}, who suggested that in ETGs the molecular gas mass drops 200\,Myr after the starburst, interpreting this as evidence that low-luminosity AGN are responsible for the destruction of the gas reservoir. We note that their sample is not directly comparable to ours as their ETGs are classified spectroscopically as blue-sequence (normal star-forming) galaxies and not PSBs, and therefore there is no well defined age for the stellar population as is the case for PSB galaxies. 

The presence of substantial molecular gas reservoirs in the 600\,Myr old PSBs suggests that these galaxies are still capable of forming stars and are not entering the red sequence. The oldest PSB (PSB11) has a gas-to-stellar mass ratio of 0.02, still significantly higher than ETGs found in the literature. If low redshift PSBs ultimately lead to red sequence galaxies, these results imply a transition timescale of more than 1\,Gyr. We can conclude that the PSBs are not transitioning to the red sequence any more rapidly than the Milky Way (MW) or any other local nearby spiral \citep{Leroy13}.

\subsubsection{Dust-to-stellar mass ratio}

The median dust-to-stellar mass ratio (\mdms) of the PSBs is $1.3\times{10}^{-3}$ with a range of $(0.3-10.0)\times{10}^{-3}$, similar to dusty galaxies in the local Universe \citep{DJBSmith12}.
In Figure~\ref{fig:ISMcontent} (b) we compare the dust-to-stellar mass ratios of the PSBs to those of other galaxies, as a function of starburst age. We observe a significant decrease in the dust-to-stellar mass ratio of the PSBs with
starburst age (Spearman coefficient $\rho=-0.71$, $\sigma=0.01$). The
dust-to-stellar mass ratios of the young PSBs are slightly higher than those of
spiral galaxies in the H-ATLAS survey which have a mean \mdms\, of $2.1\times{10}^{-3}$, and have values similar to those of ultra-luminous infrared galaxies (ULIRGs) and high-redshift submillimetre galaxies (SMGs) \citep{dC10b, Rowlands14a}. Intermediate and older PSBs have \mdms\, values
similar to those of dusty ETGs \citep{Rowlands12, Agius13}, and are higher than
those of non-detected ETGs in H-ATLAS. This shows that the older PSBs still harbour a cold ISM. The decrease in dust content with starburst age suggests that the dust in PSBs is being destroyed in the ISM and is no longer balanced by stellar dust production. 
The similarity of the \mdms\, of older PSBs and dusty ETGs suggests a possible connection, which should be investigated through further spectroscopic and morphological analyses.

\subsubsection{Gas-to-dust ratio}
\label{sec:GastoDust}

In Figure~\ref{fig:ISMcontent} (c) we show the gas-to-dust ratios (G/D) of the PSBs, as function of starburst age. The PSBs have a large range of gas-to-dust ratios, with values of $40-380$, with a median of 140. This is similar to the molecular gas-to-dust ratios of high metallicity ($12+$log(O/H)$>8.5$) nearby galaxies in the {\sc KINGFISH} sample, which have a mean of G/D of 130 (assuming a MW CO to H$_2$ conversion factor; \citealp{Remy-Ruyer14}).
In Figure~\ref{fig:ISMcontent} (c) we observe no significant correlation of gas-to-dust ratio with starburst age (Spearman coefficient $\rho=0.35$, $\sigma=0.33$). The weakness of the correlation is largely caused by one source (PSB11). Excluding this source from the correlation analysis results in a Spearman coefficient of 0.72, indicating a significant ($\sigma=0.03$) positive correlation between gas-to-dust ratio and starburst age. 
Typically the young ($<100$\,Myr) PSBs have low gas-to-dust ratios of $<100$ which is lower than normal spirals, and is more similar to that of high redshift SMGs \citep{Kovacs06, Swinbank14}. The majority of PSBs ($>100$\,Myr) have a molecular gas-to-dust ratio consistent with that of normal star-forming galaxies in the nearby Universe.

\subsubsection{Chemical evolution models}

We now compare our observations to the one-zone chemical evolution models of \citet{ME03} and \citet{Rowlands14b} to investigate the evolution of the dust and gas in PSBs. 
By relaxing the instantaneous recycling approximation to account for the lifetimes of stars of different masses, the model tracks the build-up of heavy elements over time produced by low-intermediate mass stars (LIMS) and supernovae (SNe), where some fraction of the heavy elements will condense into dust. Given an input SFH, gas is converted into stars over time, assuming a \citet{Chabrier03} initial mass function. To model the PSBs we consider a closed box model, assuming no inflow or outflow of gas or metals. This is justified during the recent evolution of the PSBs by the observed lack of change in gas-to-stellar mass ratio over time, implying that recent strong outflows are unlikely. 
The initial gas mass is set at $1.0\times 10^{10}\,\rm M_{\odot}$, and at the end of the SFH $\sim85$\% of the total galaxy mass ends up in stars and $\sim15$\% is in gas, in agreement with our observations. By design, the final stellar masses in the chemical evolution models are in close agreement with the observed stellar masses (mean \mstar$=1\times 10^{10}\,\rm M_{\odot}$) derived from the SED fitting. We adopt a constant SFH of 2\,M$_{\odot}$yr$^{-1}$ from 0--4.9\,Gyr (assuming arbitrarily that our PSBs formed at $z\sim1$), followed by a starburst of 10\,M$_{\odot}$yr$^{-1}$ of 100\,Myr duration. The starburst produces $\sim 10\%$ of the stellar mass in the galaxy, in agreement with observations \citep{Wild07, Wild09}.
At $t=5.0$\,Gyr the SFH is parameterised by an exponentially declining SFR of the form exp$-(t/ \tau)$, with  $\tau=300$\,Myr, which matches the measured SFRs of the PSBs as a function of starburst age \citep{Wild10}. 

In Figure~\ref{fig:ISMcontent} (a) the overplotted lines show that the choice of these initial conditions and SFH allow us to match the evolution of the gas-to-stellar mass with starburst age. However, we find that dust produced only by LIMS is not sufficient to match the \mdms\, and gas-to-dust ratio for the young PSBs (thin black line in Figure~\ref{fig:ISMcontent} (b) and (c)). The large dust masses in these galaxies indicates that rapid dust production (most likely from supernovae) is necessary (thick black line). 
The negative relation between \mdms\, and starburst age indicates that dust is lost from the system over time, either via outflows or dust destruction by supernova shocks or AGN feedback. The closed box model with no dust destruction results in an almost flat evolution of \mdms\, with starburst age, regardless of the dust source. We tested that reasonable variations in the assumed SFH of the PSBs are unable to reproduce the observed trends. In order to reproduce the decrease in \mdms\, with time, we require a model in which moderately strong dust destruction (m$_{\rm ISM}=500$M$_{\odot}$ of ISM cleared of dust per SN explosion) occurs for a short time after the starburst (here assumed to be from 0--100\,Myr following the starburst), after which the dust destruction is weaker (m$_{\rm ISM}=200$M$_{\odot}$). 
We note that when dust destruction is included, our models require very efficient dust production, such that all of the metals produced in supernovae are incorporated into dust. This may not be physical, therefore grain growth in the ISM may also play an important role in the dust budget \citep{Rowlands14b}. We note that dust destruction much stronger than that assumed here, e.g. m$_{\rm ISM}=1000$M$_{\odot}$, as is often assumed in chemical evolution models \citep{Dwek07}, results in a rapid increase in gas-to-dust ratio and a final ratio that is inconsistent with that observed in PSBs older than 100\,Myr. One way out of this is to invoke subsequent rapid grain growth in the ISM, but such complex models are not justified by the data presented here.

\subsection{Dust temperature}

Dust is primarily heated by UV--optical photons from stars, although AGN heating can also contribute in the central regions of massive galaxies \citep{Sauvage_Thuan94}. Dust then re-radiates the absorbed light in the FIR--submillimetre, with young stars thought to be responsible for heating warm ($>60$\,K) dust \citep[e.g.][]{Sauvage90, Xu90, Popescu00, Misiriotis01, Bendo10}. Colder dust is thought to be heated by the diffuse interstellar radiation field (ISRF; \citealp{RRC89, XuHelou96, Stevens05, Komugi11, Boquien11b}), which originates from photons from old stars \citep{Bendo10, Boquien11b, Groves12} and also young stars where photons leak out of the birth clouds \citep{Kirkpatrick14}. 

\begin{figure}
\begin{minipage}[t]{0.48\textwidth}
\begin{center}
\includegraphics[width=1.\textwidth]{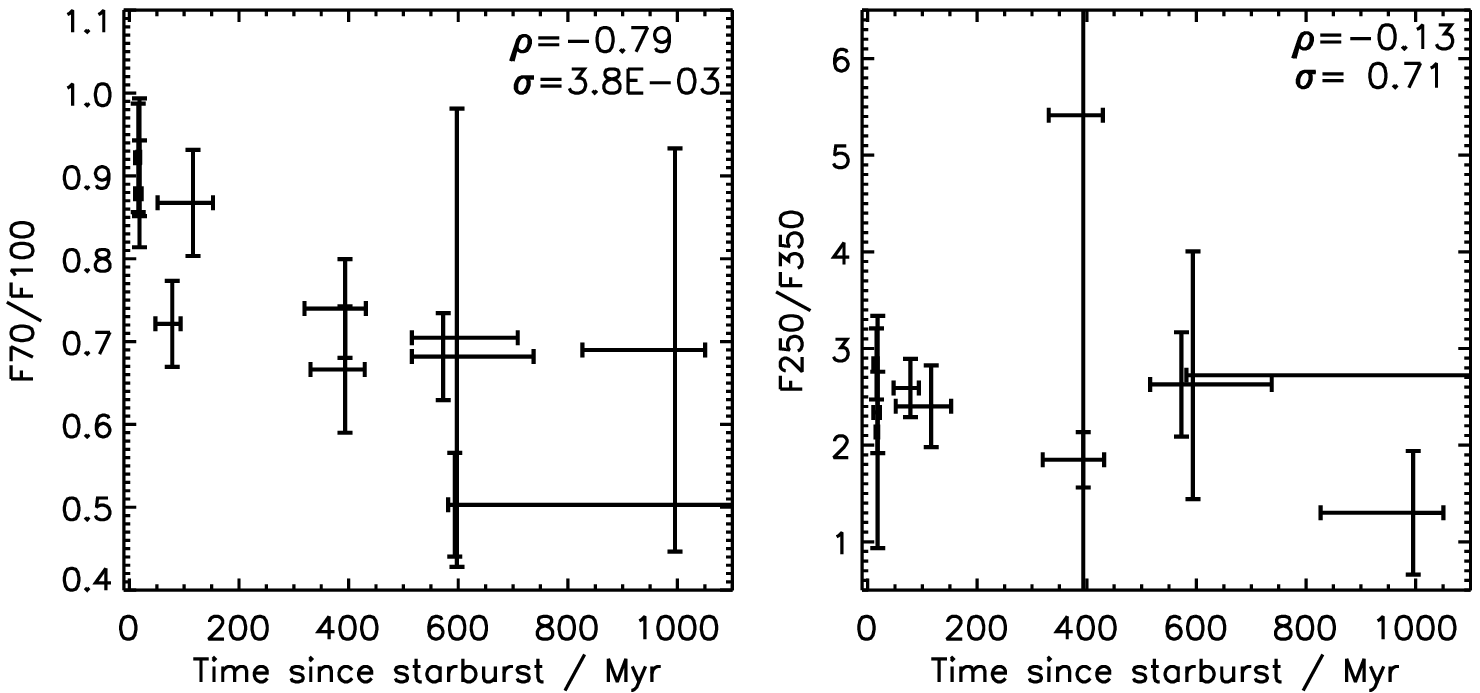}
\includegraphics[width=1.\textwidth]{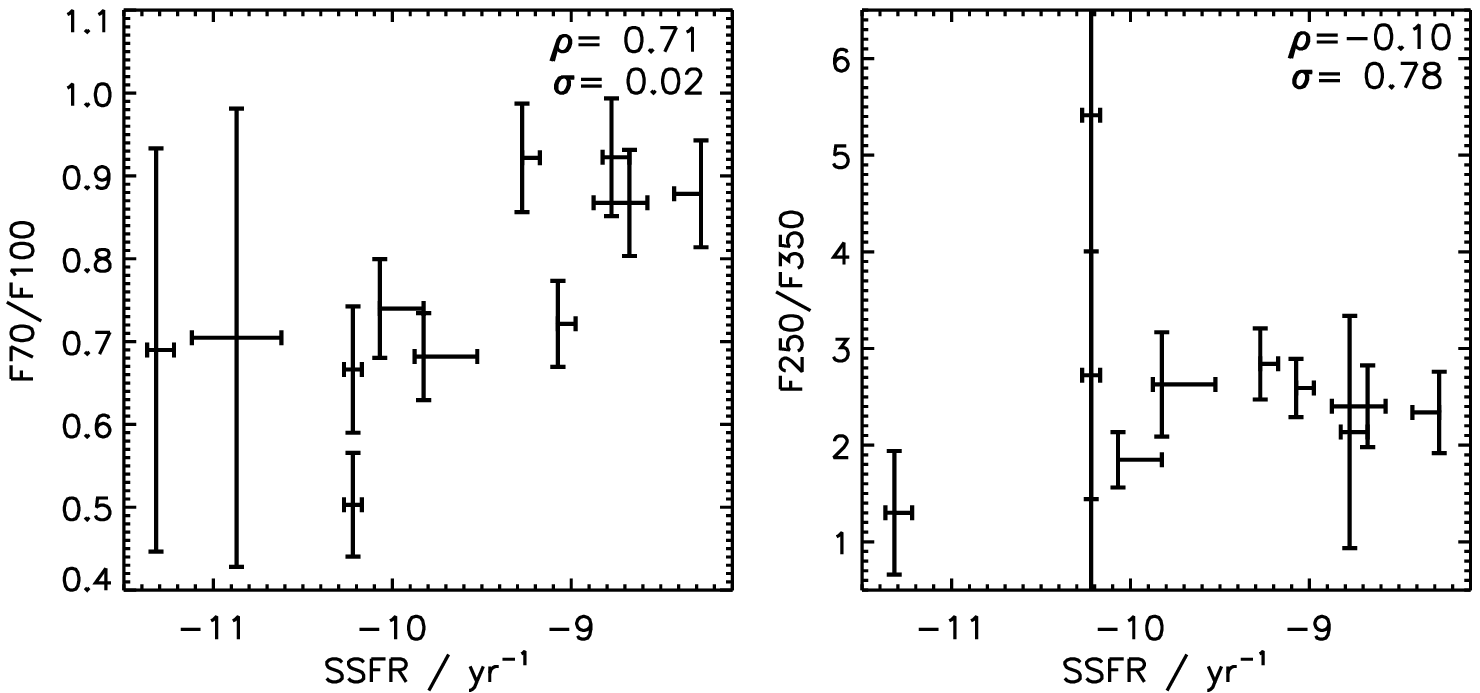}
\end{center}
\end{minipage}
\caption{FIR colours $f70/f100$ (warm dust) and $f250/f350$ (Rayleigh-Jeans slope) as a function of starburst age (upper panels) and SSFR (lower panels). Values in the right of each plot indicate the Spearman rank coefficient ($\rho$) and the two-sided significance of the deviation of the correlation from zero, or the null hypothesis ($\sigma$, a low value indicates a significant correlation). 
The observed values correspond to a mean effective cold dust temperature dropping from 28\,K-17\,K, assuming a single modified blackbody fit (see text for further details).}
\label{fig:FIR_colour}
\end{figure}

In Figure~\ref{fig:FIR_colour} we investigate the variation in the effective dust temperature as a function of starburst age. Instead of plotting the model dependent temperatures output by the {\sc MAGPHYS} code, we use the colour index $f70/f100$ as a proxy for the presence of warm dust and $f250/f350$ as a proxy for the slope of the Raleigh-Jeans tail of the SED. The observed values correspond to a mean effective cold dust temperature of 28\,K for the young ($<200$\,Myr) PSBs, dropping to a mean of 17\,K for the older ($>550$\,Myr) PSBs, using a single modified blackbody fit to the $\geq 100\mu$m fluxes (a second component is required to fit warm dust emitting at $70\mu$m), We assume a dust emissivity $\beta=2$ and dust mass absorption coefficient $\kappa_{850}=0.077\,\rm{m}^2 \,\rm{kg}^{-1}$. 
Fitting a two-component modified blackbody to the $\geq 70\mu$m fluxes leads to the same observed trends. Fixing the warm dust temperature at 45\,K \citep{Bourne13}, the cold dust temperatures from the two-component fit decrease from a mean of 26 to 16\,K for the young and old PSBs, respectively.  To test if these drops in temperature are significant, we measured the standard deviation of the temperatures for 500 Monte-Carlo realisations. We randomly perturbed each of the FIR fluxes by an amount centred on a Gaussian probability distribution of width $\sigma$ equal to the flux density uncertainty described in Section~\ref{sec:Herschel}. For each realisation we then refit the one- or two-component modified blackbody. The standard deviations of the fitted temperatures for each of the PSBs ranges from $0.5-2.2$\,K (for a one-component modified blackbody), and $0.8-2.8$\,K (for a two-component modified blackbody), which is less than the observed drop in temperature with starburst age.

We therefore conclude that there is a clear trend of decreasing effective dust temperature with increasing time since the
starburst ended (Spearman rank coefficient $\rho=-0.79$, $\sigma=4\times10^{-3}$). We find no correlation (Spearman rank coefficient $\rho=-0.13$, $\sigma=0.71$)
between $f250/f350$ and starburst age, indicating that the Raleigh-Jeans slope does not change with starburst age.

Since SSFR decreases over time in our sequence of PSBs, there is also a clear correlation of dust temperature ($f70/f100$) with SSFR (Figure~\ref{fig:FIR_colour}, lower panels; $\rho=0.71$, $\sigma=0.02$). We find no correlation between the slope of the Rayleigh-Jeans side of the SED ($f250/f350$) and SSFR ($\rho=-0.10$, $\sigma=0.78$). Our results are in agreement with \citet{Boquien11b} and \citet{Boselli10,Boselli12} who found a correlation between birthrate parameter ($b$, equivalent to SSFR) and effective dust temperature, and no correlation of $b$ with $f250/f350$ in spiral galaxies. 
\citet{Cortese14} also found that the effective dust temperature (i.e. the peak of the SED) correlates with SSFR and spatially with young stars (see also \citealp{Kirkpatrick14}). 

The same trend of dust temperature is also seen in mergers, where the warm dust mass increases with interaction stage \citep{Xilouris04, Hwang11, Lanz13}. Local ULIRGs (merger induced starbursts) are also found to have warmer dust temperatures compared to normal spirals \citep{Clements10}.

The simplest explanation for the decrease in dust temperature with time since the starburst is that the strength of the ISRF decreases over time as star-formation activity in the PSBs declines. This is also observable as a trend with SSFR, because SSFR correlates with the ratio of ionising (from young stars) to non-ionising radiation (from old stars) and therefore effectively measures the strength of the ISRF. Dust grains in young PSBs are exposed to a more energetic radiation field from UV emission from young stars, which results in higher dust temperatures. An additional decline in temperature may occur if the dust becomes more diffusely distributed as the PSB ages. Simulations of high-redshift mergers suggest that the increase in dust temperature during a starburst is caused by a combination of the dust being located in a smaller volume, alongside the dust mass decreasing during the starburst \citep{Hayward11}, allowing the dust to be heated to a higher temperature. However, the decrease in dust mass found in the simulations of starbursts is contrary to our observations, where we find our youngest PSBs to have the highest dust mass fractions. 
In summary, the decrease in dust temperature is consistent with a decline in the heating contribution from young stars as the PSB ages. We additionally note that there is no evidence of dust heating at late times by an AGN.

\subsection{Line ratios}

\begin{figure}
\includegraphics[width=0.48\textwidth]{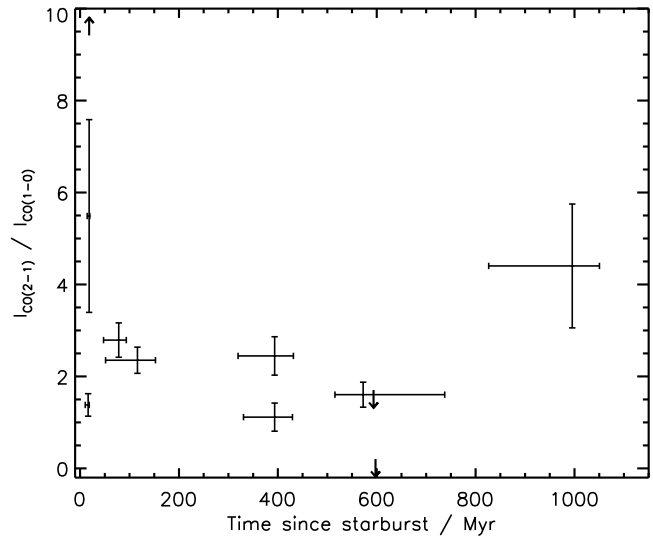}
\caption{Ratio of the observed 12CO(2-1) to 12CO(1-0) line intensities (no aperture correction applied) as a function of starburst age. The wide range of line ratios indicates that the CO emission does not fill each beam uniformly. There is no trend of line ratio with starburst age.}
\label{fig:R21}
\end{figure}

The ratio R21 of 12CO(2-1) to 12CO(1-0) line intensity depends on the optical depth, excitation temperature and spatial distribution\footnote{The measured brightness temperature (a specific intensity) is averaged over the beam area, and so the ratio of the brightness temperatures at different frequencies is determined partly by beam dilution.} of the gas. As our observations consist of one single pointing (with beam sizes of $\sim22$\arcsec and $\sim11$\arcsec for 12CO(1-0) and 12CO(2-1), respectively; i.e. a factor of four difference in area) we cannot disentangle the effect of gas distribution and physical properties. A value of R21$\sim4$ indicates that the CO is compact compared to the beam size \citep{Young11}. Assuming both lines are optically thick and have the same excitation temperature, then R21$\sim1$ indicates that the CO emission uniformly fills both beams.
In Fig.~\ref{fig:R21} we plot the integrated intensities of the CO(2-1) and CO(1-0) lines (without any aperture corrections) as a function of starburst age. We do not attempt to account for beam dilution as we do not have a good estimate of the size of the CO emitting region, leading to a large uncertainty in the intrinsic value of R21. We find that most of the PSBs have values of R21 between 1 and 4, suggesting that the CO does not fill both CO beams uniformly. Furthermore, we do not see a correlation of R21 with starburst age. This indicates that either the distribution and excitation of molecular gas does not vary with starburst age, or that the excitation of the gas is decreasing as the CO distribution becomes more compact. It is likely is that PSBs have a wide range of global excitation conditions, as well as variations within each galaxy.
Further investigation of the excitation temperatures of PSBs will require spatially resolved measurements of these sources.

\subsection{Where do PSBs lie on the SFR-\mstar\,relation?}
\begin{figure}
\includegraphics[width=0.48\textwidth]{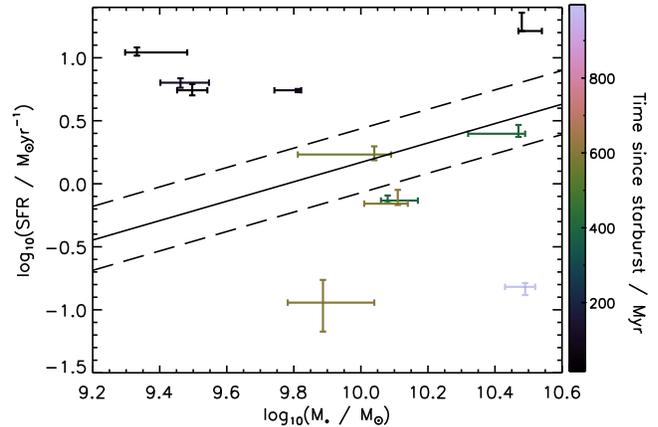}
\caption{Comparison of the total SFR and the stellar mass of the PSBs (with points colour coded by starburst age) to the median location of blue star-forming galaxies in the local Universe ($0.015<z<0.1$) from \citet{Elbaz07} (solid black line). The dashed lines indicate the 68\% dispersion in the median of the local relation between SFR and \mstar.}
\label{fig:MainSequence}
\end{figure}

Clues to the stellar mass build-up of galaxies can be found via the relatively tight correlation between stellar mass and SFR \citep[e.g.][]{Brinchmann04, Noeske07, Elbaz07, Daddi07, Rodighiero10, Wuyts11b, Whitaker12b}. In Figure~\ref{fig:MainSequence} we compare the location of the PSBs to those of blue star-forming galaxies in the local Universe ($0.015<z<0.1$) from \citet{Elbaz07}. The SFRs of the galaxies in \citet{Elbaz07} were taken from \citet{Brinchmann04} and utilise the dust-corrected H$\alpha$ luminosity (corrected to total SFR), and the stellar masses are from \citet{Kauffmann03a} and use the \citet{BC03} stellar population models. We therefore plot the equivalent observations for the PSBs. The young PSBs lie significantly above the local SFR-\mstar\,relation, with the intermediate-age PSBs coincident with normal star-forming galaxies. Two PSBs lie significantly below the location of blue star-forming galaxies and are consistent with those of $0.02<z<0.05$ green-valley galaxies \citep{Schawinski14}.

In the last few years, significant attention has been given to galaxies 
which lie below the SFR--\mstar\,relation \citep[e.g.][Eales et al. in prep]{Wuyts11b, Whitaker12b}, which are assumed to be ``quenching" (i.e. switching off their star formation). Given the extra information provided by our detailed SFHs, we have a different picture of where quenching galaxies lie relative to the SFR-\mstar\,relation: galaxies which are ``quenching" can lie on, above and below the location of star-forming galaxies. Furthermore, we have shown in this paper that the older PSBs are not fully ``quenched", with considerable cold gas supplies remaining. The limited information provided by instantaneous SFR and \mstar\,does not give a full picture of the physical mechanisms of quenching. Caution should therefore be exercised when classifying galaxies on the SFR--\mstar\,relation as star-forming galaxies which have remained, and will remain, on the relation for an extended period of time.

\subsection{Environment}
The large-scale environment of galaxies affects their gas supply. In clusters, the supply of gas can be terminated via interaction of the galaxy ISM with the intracluster medium, which heats or strips gas from the galaxy. In less dense regions such as the field, gas may be readily accreted from the intergalactic medium, although galaxy interactions in small groups can alter gas supplies by triggering a starburst.

We examine the environment of the PSBs in our sample by matching to the SDSS DR7 galaxy group catalogue of \citet{Yang07}. The group catalogue utilises a halo-based group finder to identify galaxies which reside in the same dark matter halo. We use the group catalogue based on SDSS model magnitudes and redshifts from both SDSS and the 2dF Galaxy Redshift Survey. Halo masses are assigned by rank ordering the groups by stellar mass, although using masses assigned by rank ordering by luminosity would not change our results. We find 10/11 PSBs are identified in the group catalogue, 6 of which are isolated galaxies, and four PSBs are in groups with 12--59 members. The halo masses of the groups range from $10^{11.9}-10^{14.6}$M$_\odot$. Our PSBs therefore reside in a variety of galaxy environments, in agreement with previous studies \citep{DresslerGunn83, Zabludoff96, Quintero04, Blake04, Balogh05, Hogg06, Poggianti09, Yan09}.

\section{Discussion and conclusions}
\label{sec:conc}

It is often stated in the literature that PSBs could be a channel for the growth of the red sequence, through migration from the blue cloud to the red sequence over a time-scale of 100 Myr to 1Gyr. \citep{Kaviraj07, Wild09}. Transition through the PSB phase potentially represents the fastest ``quenching" mode available to galaxies \citep{Faber07, Martin07, Barro13, Barro14, Yesuf14}, requiring a dramatic starburst event which can rapidly exhaust gas supplies. Such an event may be triggered by an external interaction or merger, certainly at lower redshifts where gas mass fractions are low. Slower quenching processes may also occur through secular processes \citep[][Eales et al. in prep]{CorteseHughes09, Salim12, Fang12, Fang13, Schawinski14}, that transfer galaxies through the green-valley on longer timescales. Revealing the existence, and understanding the relative importance, of these two channels for building the red sequence is one of the most important challenges in galaxy evolution.   

We have examined the global cold gas, dust and star-formation properties of a sample of 11 PSB galaxies which have undergone a central starburst in the last 0--1\,Gyr. We derived the physical properties of the PSBs using a multi-wavelength dataset with optical fibre spectroscopy of the central 3'', total broadband photometry from the UV--FIR, and CO measurements which trace the global molecular gas reservoir. 
As our sample represents some of the most extreme starbursts in massive galaxies in the SDSS volume, we would expect them to experience the fastest quenching of star formation. However, the existence of substantial cold gas reservoirs in all the PSBs studied here suggest that the transition time for local galaxies to become truly ``red and dead" with no possibility for subsequent star formation is $>>$1\,Gyr, i.e. not particularly fast. The significant cold gas supplies in the recently ``quenched" galaxies, makes it unclear as to whether the galaxies will continue evolving onto the red sequence, or remain in the ``green valley" for an extended period of time (see also \citealt{French15}).

AGN feedback is often proposed as a mechanism for fully halting star formation following a starburst (either through gas outflows or heating, \citealp[e.g.][]{DiMatteo05, Croton06, Schawinski07, Schawinski09, Kaviraj07}). The lack of change in gas-to-stellar mass ratio with increasing starburst age shows that the bulk of the cold ISM remaining after a starburst is not expelled by stellar winds or AGN feedback.  Similarly, the steady decrease in dust temperature indicates that the global ISM is not heated by AGN feedback at later times. For the sample of objects studied in this paper, WHC10 found that the peak in black hole accretion rate occurred $\sim$250\,Myr following the starburst and therefore could not be contributing to the initial rapid decline in star formation observed $\sim10^7$ years following the starburst \citep[see also][for radio-loud AGN]{Shin11}. In this paper we show that there is no significant alteration in the physical conditions of the cold ISM that may be attributed to the late time triggering of an AGN.

Our sample provides a new way to study the alteration in global ISM conditions following a burst of star formation. We observe a steady decline in star formation efficiency with starburst age, consistent with a natural decrease in the SFR following consumption of the dense gas reservoir (not traced by our CO observations) by the starburst. A similar steady decrease is observed in the effective dust temperature, which is likely due to a decrease in the ISRF strength as the starburst ages. In the young PSBs we observe a high dust-to-stellar mass ratio, which declines with increasing starburst age. 

Using chemical evolution modelling we show that rapid and efficient dust production must be important during the starburst, and dust must be formed by both low-intermediate mass stars and SNe in order to match the dust-to-stellar mass ratio and the gas-to-dust mass ratio. The observed decrease of the dust-to-stellar mass ratio with starburst age cannot be modelled only by low rates of dust production due to declining star-formation activity. Instead, models must include moderately strong dust destruction due to supernova shocks which weakens with starburst age, possibly due to the declining supernova rate or changes in density of the ISM. 

The gas-to-stellar mass ratio ($\sim0.2$), dust-to-stellar mass ratio ($1\times{10}^{-3}$) and gas-to-dust mass ratio (100--300) of the oldest PSBs in our sample with ages $\sim0.5-1$\,Gyr are consistent with those of normal spiral galaxies and dust/gas-rich ETGs, and do not resemble those of normal ETGs or red sequence galaxies. The similarity of the morphology and physical properties (gas-to-stellar mass ratio, dust-to-stellar mass ratio) of the older PSBs to dusty/molecular gas-rich ETGs suggests that these populations may be related, or may have similar endpoints in cold ISM content after a starburst. Further comparisons may reveal that these two populations are intimately linked. 

As the central starburst has been identified through 3" fibre spectroscopy, it is possible that the lack of change in the molecular gas-to-stellar mass ratio may be because the central starburst does not affect the global properties of the galaxy. However, we showed that the global SFR estimated by SED fitting tracks the central SFR as the starburst ages with only a small offset in some objects. It therefore seems unlikely that we will find a significant offset between central and global gas reservoirs. Significant aperture effects certainly do exist in some local SDSS-selected PSBs, for example \citet{Pracy14} found that star formation is ongoing outside of the PSB core in two very low redshift HI-rich galaxies. It is possible that galaxies that undergo a strong global (rather than central) starburst may be more efficient at completely exhausting or expelling their cold ISM and be a more effective channel by which the red-sequence is built. It is also possible that such intense events may be more common at high redshift \citep{LeBorgne06, Wild09, Wild14}. Such galaxies can now be selected in large multi-wavelength surveys via their UV--near-infrared colours as demonstrated in \citet{Wild14}; these truly global PSBs will be the topic of future investigations.

Recent work on our own Galaxy, the Milky Way (MW), has found that it lies in the ``green valley" \citep{Mutch11}, and its star formation has been roughly constant over the last 3 Gyr \citep{Hernandez00}, and perhaps
for the last $\sim$8 Gyr \citep{Haywood13, Snaith14}. A constant star
formation rate suggests that the MW must have been in the green valley
for a substantial time (last few Gyrs). However, the time resolution
of the SFH of the the Milky Way is $\sim1$\,Gyr, therefore it is quite possible that bursts of star formation have occurred from time to time, as observed in the objects studied in this paper. It is important to note that despite the status of the MW as a green valley galaxy, it is considered to be unlikely that it is rapidly heading to the red sequence as its gas depletion time is very similar to other local spiral galaxies \citep{Leroy13} and to the older PSBs studied here ($>300$\,Myr; Fig. \ref{fig:Tdep}). Thus, the MW would fit into a picture where starbursts in local galaxies are followed by a return to pre-disruption normality, although possibly in a slightly earlier-type system, and may lie in the green valley for a few Gyr or longer.

In this paper we conclude that, while the SFR of PSB galaxies in the local Universe steadily declines with time following the starburst, substantial cold gas reservoirs remain to fuel future star formation, at least up to 0.5-1\,Gyr following the starburst. It therefore seems unlikely that massive galaxies in the local Universe which have undergone a strong starburst will contribute to an increase in the number of red sequence galaxies on even a 1\,Gyr timescale, and they may remain in the ``green valley" for an extended period of time. Morphological analyses of PSB time sequences will help to understand whether the galaxies become increasingly ``early-type" following such starburst episodes. Further observations of the cold ISM in a larger number of older ($\gtrsim$1\,Gyr) PSB galaxies are required in order to track their late-time evolution, and reveal whether the decline in star formation continues and whether they enter the red sequence despite their residual cold gas supplies, perhaps as gas-rich ETGs. Ultimately, studies of galaxies in the local Universe are limited to revealing only what is happening currently. In order to understand the origin of the present day galaxy bimodality, such observations will have to be repeated at higher redshifts, where present day red sequence/elliptical galaxies are forming.

\section*{acknowledgements}
We thank the referee for useful suggestions which improved the clarity of the paper,
and the IRAM duty astronomer Hans Ungerecht for his help and advice during the observations. V.~W. and K.~R. acknowledge support from the European Research Council Starting Grant SEDmorph (P.I. V.~Wild). 
This publication has made use of code written by James R. A. Davenport, the IDL Astronomy Library \citep{Landsman93} and David Fanning's Coyote IDL library (http://www.idlcoyote.com/).
{\it Herschel} is an ESA space observatory with science instruments provided by European-led Principal Investigator consortia and with important participation from NASA. This publication makes use of data products from the Wide-field Infrared Survey Explorer, which is a joint project of the University of California, Los Angeles, and the Jet Propulsion Laboratory/California Institute of Technology, funded by the National Aeronautics and Space Administration. 
Funding for the SDSS and SDSS-II has been provided by the Alfred P. Sloan Foundation, the Participating Institutions, the National Science Foundation, the U.S. Department of Energy, the National Aeronautics and Space Administration, the Japanese Monbukagakusho, the Max Planck Society, and the Higher Education Funding Council for England. The SDSS Web Site is http://www.sdss.org/.
The SDSS is managed by the Astrophysical Research Consortium for the Participating Institutions. The Participating Institutions are the American Museum of Natural History, Astrophysical Institute Potsdam, University of Basel, University of Cambridge, Case Western Reserve University, University of Chicago, Drexel University, Fermilab, the Institute for Advanced Study, the Japan Participation Group, Johns Hopkins University, the Joint Institute for Nuclear Astrophysics, the Kavli Institute for Particle Astrophysics and Cosmology, the Korean Scientist Group, the Chinese Academy of Sciences (LAMOST), Los Alamos National Laboratory, the Max-Planck-Institute for Astronomy (MPIA), the Max-Planck-Institute for Astrophysics (MPA), New Mexico State University, Ohio State University, University of Pittsburgh, University of Portsmouth, Princeton University, the United States Naval Observatory, and the University of Washington.


\bibliographystyle{mn2e}
\bibliography{refs_all}

\begin{appendix}

\section{Post-starburst galaxy selection}
\label{sec:PSB_classical_selection}
Samples of PSB galaxies are traditionally selected to lack emission lines which trace star formation (e.g. H$\alpha$, [OII]), and exhibit deep Balmer absorption lines which signify a dominant A/F stellar population. We briefly examine which of our PSBs selected via our PCA technique would be selected as PSBs using the criteria of \citet{Goto03, Goto05, Goto07}, who require H$\alpha$ equivalent width (EW)$>-2.5$\AA, H$\delta$ EW$>5$\AA, [OII] EW$>-3$\AA (where absorption is positive EW and emission is negative EW).

\begin{figure}
\includegraphics[width=0.48\textwidth]{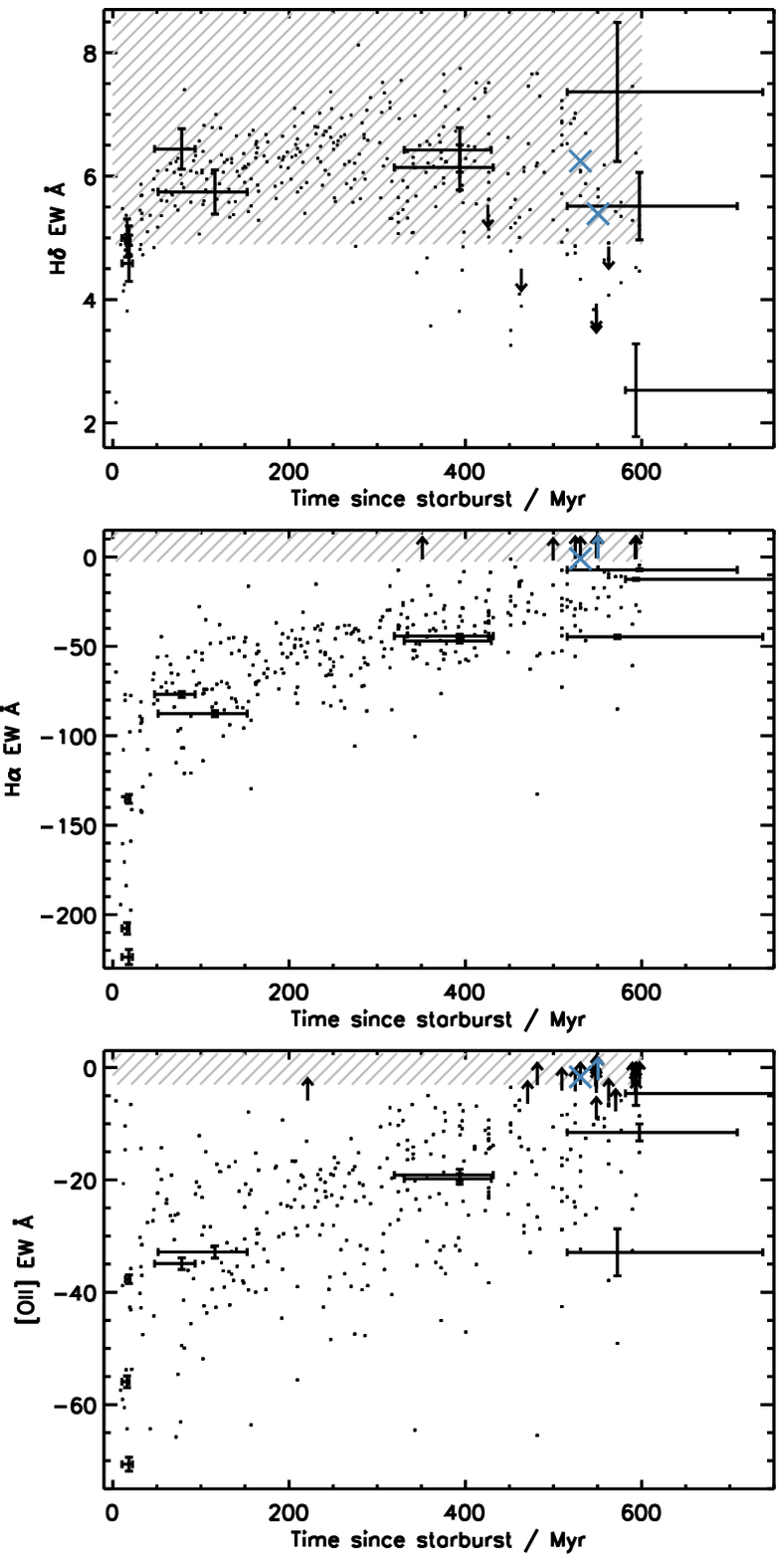}
\caption{Equivalent widths of the H$\delta_A$, H$\alpha$ and [OII] emission lines as a function of starburst age for the SDSS bulge-dominated PSB parent sample (dots) and the 11 PSBs studied in this paper (thick points with error bars). The grey areas denote the equivalent width at which a galaxy would be identified as a PSB galaxy by \citet{Goto03}. The blue crosses indicate galaxies which meet all three criteria required by \citet{Goto03} to be classified as a PSB. Upper limits at the $3\sigma$ level are shown for galaxies with SNR$<3$, sources with blue upper limits have SNR$<3$ in either H$\alpha$ or [OII] EW and meet the \citet{Goto03} criteria at the $3\sigma$ level.}
\label{fig:LineProperties}
\end{figure}

In Figure \ref{fig:LineProperties} we show the line EWs as a function of starburst age for the 11 PSBs studied in this sample and the SDSS bulge-dominated PSB parent sample. The line equivalent widths are from the MPA-JHU catalogue\footnote{http://www.mpa-garching.mpg.de/SDSS/DR7/. We apply the appropriate scaling to the EW errors as recommended in the MPA-JHU documentation.}. Of the 11 PSBs studied in this paper, none would meet the \citet{Goto03} criteria due to the presence of H$\alpha$ and [OII] emission. Of the parent sample of PSBs with starburst age $<600$\,Myr and measurements of all three lines, only 5/377 meet the criteria, although 4 of these sources have SNR$<3$ in either H$\alpha$ or [OII] EW\footnote{Note that \citet{Goto03} do not apply a SNR cut to their EW measurements.}. Of these four sources, only one PSB meets the \citet{Goto03} criteria at the $3\sigma$ level. It can be seen from Figure \ref{fig:LineProperties} that selection criteria requiring no emission only select PSBs with ages older than 400\,Myr. This is primarily due to the relatively slow decline in SFR following a starburst, although additionally it excludes all narrow-line AGN which are more prevalent in PSBs than other galaxies \citep{Yan06, Wild10}. Samples where an emission line cut has been applied cannot be used to study the evolution of starbursts into PSBs.

\section{Spectra and images}
\label{sec:spec}

\begin{figure*}
\includegraphics[width=0.85\textwidth]{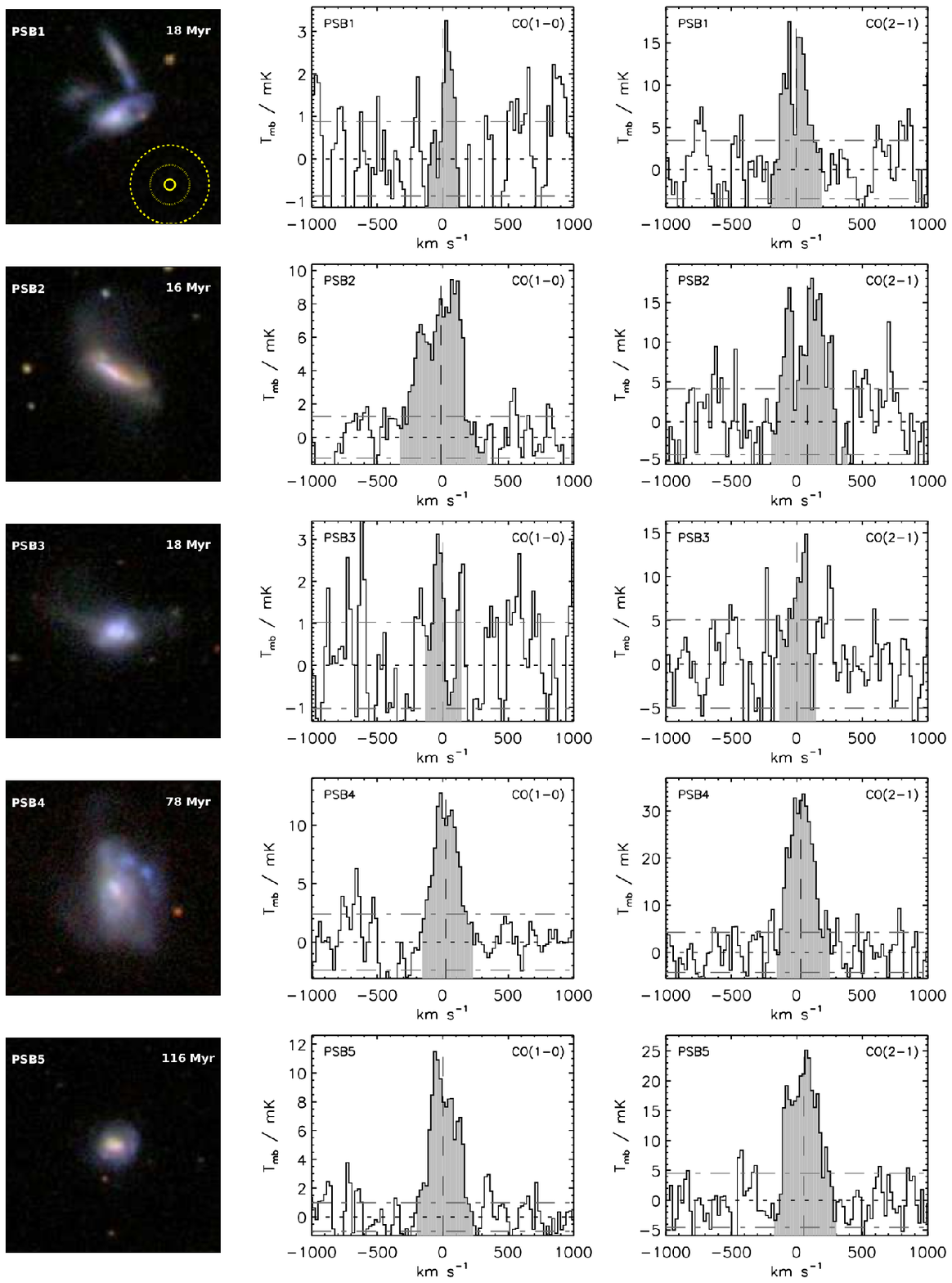}
\caption{Left: SDSS $gri$ cutouts of the PSBs, cutouts are 60\arcsec on a side. In the first image we show the extent of the 3\arcsec SDSS fibre (solid), the CO(1-0) beam (dashed) and the CO(2-1) beam (dotted). The CO 12CO(1-0) (centre) and 12CO(2-1) (right) spectra in units of main beam brightness temperature for the PSBs. The x-axis indicates the velocity offset of the line corrected to the systemic velocity of the galaxy from optical spectra. The raw CO(1-0) and CO(2-1) data has been smoothed with a boxcar of 10 and 15 velocity resolution elements, respectively, and binned to 21\,km s$^{-1}$. The grey shaded area indicates the CO line, and the dashed lines indicate the $3\sigma$ scatter around the baseline.}
\label{fig:cospec}
\end{figure*}

\begin{figure*}
\includegraphics[width=0.85\textwidth]{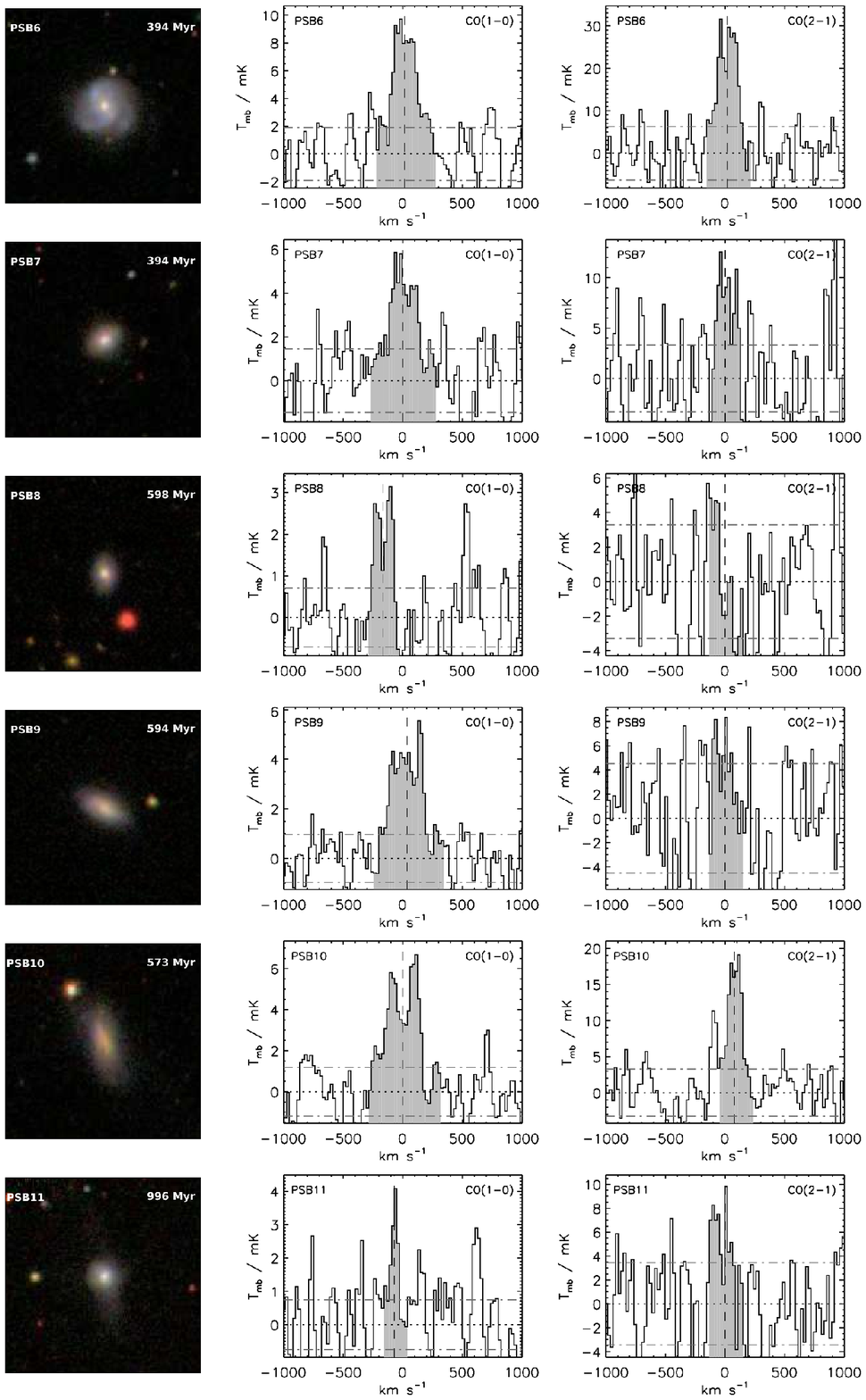}
\contcaption{}
\end{figure*}

\begin{figure*}
\includegraphics[width=0.72\textwidth]{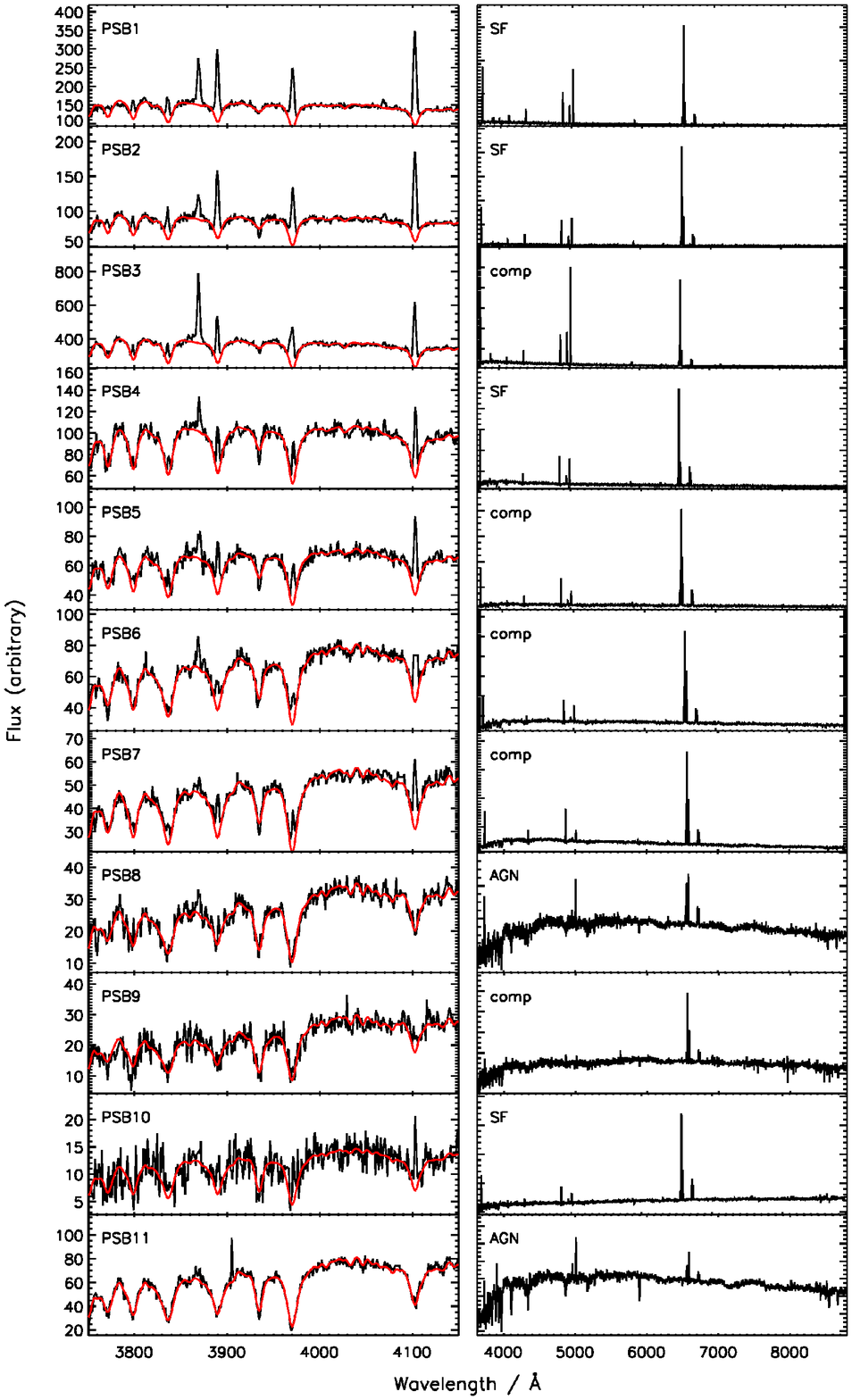}
\caption{The optical spectra of the 11 PSBs. The left
  panel shows the 4000\AA\ break region in detail, with the fitted
  eigenspectra over plotted in red (the emission lines are masked
  during the fitting). The right panel shows the full extent of the
  SDSS spectra. The spectra are ordered by starburst age: $<20$Myr
  (PSB1, 2, 3), $\sim100$Myr (PSB4 and 5), $\sim$400Myr (PSB6, 7),
  $\sim$600Myr (PSB8, 9 and 10) and $\sim$1Gyr (PSB11).}
  \label{fig:spec}
\end{figure*}

\section{SED fits}
\label{sec:SEDs}

\begin{figure*}
\includegraphics[width=0.9\textwidth]{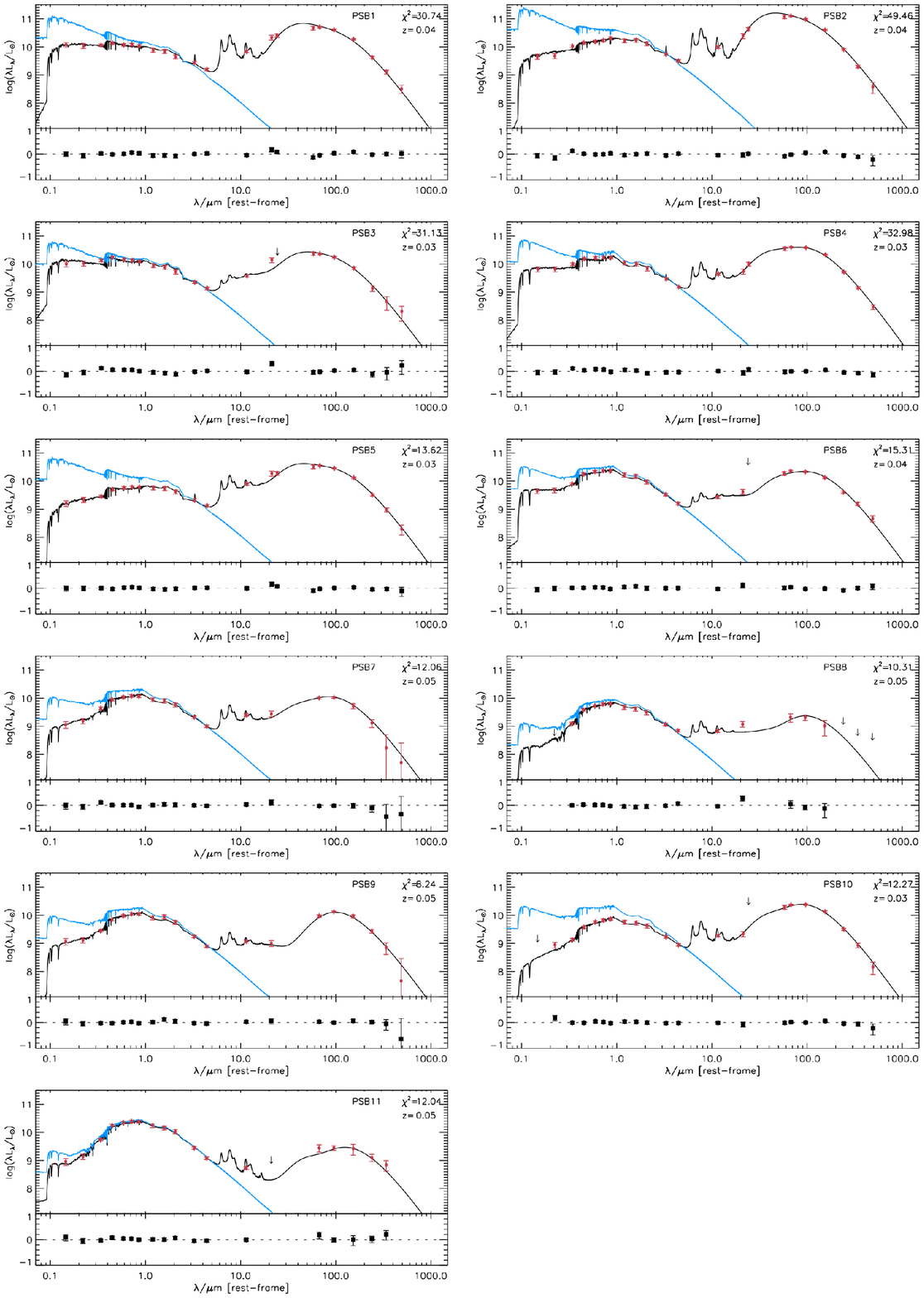}
\caption{Multi-wavelength SEDs of the 11 PSBs in our sample, with observed photometry (red points) from the rest-frame UV to the submillimetre. Upper limits are shown as arrows, and errors on the photometry are described in Section~\ref{sec:data}. The solid black line is the best-fit model SED and the solid blue line is the unattenuated optical model. The residuals of the fit are shown in the panel below each SED.}
\label{fig:SED_example}
\end{figure*}

\end{appendix}

\end{document}